\documentclass[reprint,aps,showpacs,
twocolumn,groupedaddress]{revtex4-1}
\usepackage{amsmath,amssymb,graphicx,color}
\usepackage{dcolumn}
\usepackage{bm}
\usepackage{soul} 
\usepackage{hyperref}
\usepackage{color}
\numberwithin{equation}{section}

\begin{document}

\title{Giant In-Particle Field Concentration and Fano Resonances at Light Scattering by
High-Refractive Index Particles}

\author{\firstname{Michael~I.} \surname{Tribelsky}}

\affiliation{Faculty of Physics, Lomonosov Moscow State University, Moscow, 119991, Russia}
\affiliation{Moscow State University of Information Technologies, Radioengineering and Electronics (MIREA), Moscow 119454, Russia}
\email{\mbox{tribelsky\_at\_mirea.ru}; replace \_at\_ by @}

\author{Andrey E. Miroshnichenko}
\affiliation{Nonlinear Physics Centre, Australian National University, Acton, ACT 2601, Australia}
\email[ Corresponding author:\\ ]{\mbox{andrey.miroshnichenko\_at\_anu.edu.au}; replace \_at\_ by @}

\begin{abstract}
A detailed analytical inspection of light scattering by a particle with high refractive index $m+i\kappa$ and small dissipative constant $\kappa$ is presented. We have shown that there is a dramatic difference in the behavior of the electromagnetic field within the particle (inner problem) and the scattered field outside it (outer problem). With an increase in $m$ at fix values of the other parameters, the field within the particle asymptotically converges to a periodic function of $m$. The electric and magnetic type Mie resonances of different orders overlap substantially. It may lead to a giant concentration of the electromagnetic energy within the particle. At the same time, we demonstrate that identical transformations of the solution for the outer problem allow to present each partial scattered wave as a sum of two partitions. One of them corresponds to the $m$-independent wave, scattered by a perfectly reflecting particle and plays the role of a background, while the other is associated with the excitation of a sharply-$m$-dependent resonant Mie mode. The interference of the partitions brings about a typical asymmetric Fano profile. The explicit expressions for the parameters of the Fano profile have been obtained “from the first principles” without any additional assumptions and/or fitting. In contrast to the inner problem, at an increase in m the resonant modes of the outer problem die out, and the scattered field converges to the universal, $m$-independent profile of the perfectly reflecting sphere. 
Numerical estimates of the discussed effects for a gallium phosphide particle are presented.
\end{abstract}


\date{\today}

\maketitle

\section{Introduction}
Presently the resonant light scattering by particles related to excitation of different eigenmodes attracts a great deal of attention of researchers all around the world~\cite{Zhao:MT:2009,Rybin:PRL:2009,Evlyukhin:PRB:2011,Staude:ACSN:2013,Hancu:NL:2014,Kuznetsov:NC:2014}. In addition to purely academic interest there is a broad spectrum of applications of the phenomenon in physics, chemistry, biology, medicine, data storage and processing, telecommunications, micro- and nanotechnologies, etc., see, e.g., \cite{Novotny:Book:2006,Rybin:PRB:2013}. In particular, plenty of hopes were pinned on the resonant excitation of localized and/or bulk plasmons in metal nanoparticles~\cite{Klimov:Book:2014}. Unfortunately, plasmonic resonances in such nanoparticles are usually accompanied with rather large dissipative losses, which in many cases diminish the advantages of the resonances. For this reason recently the frontier of the corresponding study has been shifted to light scattering by dielectric particles with low losses and high refractive index (HRI) $m$. In contrast to the plasmonic resonances, they exhibit the high $Q$-factor Mie resonances of both electric and magnetic types~\mbox{\cite{Evlyukhin:PRB:2010,Kuznetsov:SR:2012,Evlyukhin:NL:2012},} which bring more opportunities for wider applications in sensing, spontaneous emission enhancement, and unidirectional scattering.

Despite the fact that the exact Mie solution, describing light scattering by a sphere with an arbitrary size and material properties, is known for more than a hundred years, and the case of a sphere with HRI has been repeatedly discussed in textbooks and monographs, see, e.g.~\cite{Landau:T8:1984,Hulst::1981}, some important peculiarities of this problem have not been disclosed yet. Meanwhile, the quantitative feature of HRI brings about qualitatively new effects and paradoxes, which merely do not exist at moderate values of the refractive index, see, e.g.~\cite{Evlyukhin:PRB:2010,Tribelsky:EPL:2012,Geffrin:NC:2012}. In the present paper we produce a detailed, systematic study of light scattering by a HRI particle. Specifically, we show that the scattering may be accompanied with a \emph{giant concentration} of the electromagnetic field within the particle and reveal the nature of the Fano resonances exhibited by partial scattered waves.

It is well known that at the limit $m \rightarrow \infty$ a small (relative to the incident light wavelength) dielectric sphere scatters light as a perfectly reflecting one (PRS) ``into which neither the electric, nor the magnetic field penetrates"~\cite{Landau:T8:1984}. Then, it may be concluded that the electromagnetic field within the scattering particle should vanish at $m \rightarrow \infty$. Seemingly, the conclusion is supported by the argument that a HRI implies a high polarizability of the sphere. Then, at $m \rightarrow \infty$ from the point of view of the polarizability by an external electric field a dielectric sphere becomes equivalent to a perfectly conducting one~\cite{Landau:T8:1984}, and the electric field induced within the particle owing to its polarization by the incident light should compensate the field inducing the polarization. That is to say, the field within the particle should vanish.

In fact, the question is much more subtle, and the actual situation is far from this simple picture. The point is that the wavelength inside the particle vanishes at $m \rightarrow \infty$. Then, no matter how small the particle is, at large enough $m$ the wavelength within the particle becomes smaller than the particle size. In this case the incident wave may resonantly excite the Mie electromagnetic eigenmodes in the particle. Moreover, an unlimited growth in $m$ results in infinite cascades of these resonances. The interference of the resonant eigenmodes with the incident wave and, what is the most important, with each other gives rise to dramatic changes in the aforementioned simple scattering process. To reveal these changes is the goal of our study. To this end the full Mie theory is employed.

We show that while at $m \rightarrow \infty$ the scattered field for the outer problem does converge to the one for the PRS (no matter, whether the sphere is small, or large), the field within the particle, though it sounds paradoxical, does not have any limit at all. Such a difference between the outer and inner problems is related to the different lineshapes of the Mie resonances in the two problems. For the former an increase in $m$ makes the resonances less pronounced. For the latter in the non-dissipative limit the amplitude of the resonances increases with an increase in $m$. In the case of a finite dissipation rate (regardless how small it is), the growth of the amplitudes eventually saturates and the resonance lines become periodic functions of $m$. In both the cases the field within the particle does not tend to any fixed limit at $m \rightarrow \infty$.

It is important to stress, that the resonance lines of different orders and different origin (i.e., electric and magnetic) may overlap substantially. The mentioned peculiarities of the inner field in the vicinity of the resonances may result in a \emph{giant concentration} of the electromagnetic energy inside the particle. At realistic values of the refractive index and the proper selection of the particle radius the field inside the particle may exceed the one in the incident wave in several orders of magnitude. Such a huge field may give rise to numerous nonlinear effects. For this reason the discussed results may appear extremely important in the design and fabrication of highly-nonlinear nanostructures.

Regarding the outer problem, it is known that the scattering coefficients in this problem have the well-pronounced asymmetric Fano resonance lines. Recent publications of Rybin et al, Ref.~\cite{Rybin:OE:2013,Rybin:FTT:2014,Rybin:SciRep:2014} should be mentioned in this context. Based on the analysis of the exact Lorenz-Mie solution for a cylinder the authors of these publications have revealed that the resonant Mie scattering can be presented through infinite cascades of the Fano resonances between the narrow-line resonant Mie scattering and the non-resonant (background) scattering from the object. The analytical expressions for both the partitions have been obtained through the Maxwell boundary conditions. The numerical fit of the lineshape, resulting from the exact solution in the vicinity of the resonances, to the conventional Fano profile~\cite{Fano:PR:1961} has allowed the authors to obtain the dependence of the Fano asymmetry parameter $q$~\cite{Fano:PR:1961} on the ratio of the radius of the cylinder R to the wavelength of the incident light $x =2\pi R/\lambda$ (size parameter) in rather a broad range of its variations. They also have shown that in the inspected cases  $q(x) \sim −\cot x$. This dependence agrees with their previous results for disordered photonic crystals~\cite{Rybin:NatCom:2012}, as well as with the general expression for $q$ in terms of the phase shift of the background partition~\cite{Connerade:RPP:1988}.

Despite the study of these authors is a big step forward to understanding the essence of the Fano resonances at light scattering by a particle, they have not disclosed the physical nature of the background partition. Regarding the results obtained by the numerical fit, the great advantage of this procedure is the possibility to fit any curve with any set of the basic functions. However, precisely because of that, based on the fitting solely, one never can answer the question whether the studied profile is the Fano profile indeed, or it is \emph{just fitted} to that profile. It also remains unclear how far beyond the inspected numerical domain the obtained results could be extended, e.g., what happens with the modes with the multipolarity higher than that, examined by the authors, etc.

For this reason, a self-consistent analytical examination of the problem, connecting the parameters of the Fano profiles with the fundamental parameters of the light scattering $x$ and $m$ would be highly desirable. Such a study is produced in the present paper. Specifically, we show that the Fano profile could be obtained by identical transformations of the exact Mie solution. As a result, the exact analytical expressions for the parameters of the profile are obtained automatically. Regarding the background partition, we reveal that it is just the corresponding partial wave scattered by a PRS.

We should emphasize that the resonances discussed in the present paper have nothing to do with the well known \emph{whispering gallery modes}. These modes are associated with waves propagating along the surface of the particle, confined there by ``continuous total internal reflection"~\cite{Vahala:Nature:2003} and cannot be excited by a plane incident wave. The closest to the topic of our study is another well known phenomenon: the ripple structure in the spectrum at light scattering by droplets with $m \approx 1.5$ and related problems, see, e.g.,~\cite{Chylek:JOSA:1976,Hulst::1981,Chylek:ApplPhys:1985}. The resonances we discuss and the ripple structure, both have the same nature. However, the large value of the refractive index makes the manifestation of our resonances quite different from what is known for the ripple structure. In addition, the specific characteristics of the resonances examining in our paper, usually are not discussed in connection with the ripple structure.

Note also the very close effects discussed in Ref.~\cite{Arruda:JOSA:2010} for a magnetic particle. However, up to now materials with large  magnetic permeability at optical frequencies remain hypothetical objects, while the ones with HRI may be easily found among the common semiconductors, see below.

It is necessary to stress in this context that all the effects discussed in the present paper occur to \emph{optically thick} particles, i.e., $mx$ should be of the order of unity, or larger than that. However, an important peculiarity of the HRI particles is that this condition may hold for \emph{geometrically small} particles with $x \ll 1$, cf.~\cite{Tribelsky:EPL:2012}.

The paper is arranged as follows. In Section II we consider general properties of the Mie solution at the limit of large, purely real $m$. In Section III the lineshapes and linewidths of partial resonances for the scattered field and the field, concentrated within the particle are inspected. In Section IV the generic nature of the Fano resonances for partial modes of the scattered field is disclosed, and explicit expressions for the parameters of the Fano profile are obtained from the first principles by means of identical transformations of the exact Mie solution. In \mbox{Section V} we show that in the non-dissipative limit the entire set of the infinite number of the cascades of the resonances possesses a certain scaling and may be reduced to a universal set of lines by simple scale transformations. In Section VI, effects of finite dissipation are inspected. In Section VII the resonances at a fixed $m$ and varying size parameter are discussed and the manifestation of the resonances in a particle made of gallium phosphide is presented as an example. In Conclusion a brief summary of the obtained results is presented. In Appendix certain cumbersome but important calculations are performed.

Last, but not least. The only thing we do below is a detailed analysis of the well known Mie solution at the range of high refractive index of the scattering sphere. It would have been nothing but a mathematical exercise, if it did not reveal new unusual features of the phenomenon. It did. Let us proceed discussing these features in detail.

\section{Large refractive index limit}\label{sec:2}
The subsequent analysis is based upon the exact Mie solution describing light scattering by a spatially homogeneous sphere with an arbitrary radius $R$ and a given permittivity $\varepsilon$. The case of a cylinder and core-shell structures may be inspected in the same manner.

Since we are interested in the optical properties of a particle with low dissipation, we first focus on a discussion of the non-dissipative limit, i.e., a purely real positive refractive index $m \equiv \sqrt{\varepsilon}$. The general case of a complex refractive index along with the applicability condition for the non-dissipative limit will be produced later on.

According to the Mie solution, the scattered field is presented as an infinite series of partial multipolar contributions (dipolar, quadrupolar, etc.) of the two types: TE and TM --- the so-called electric and magnetic modes. For the sake of briefness the electric responses only are discussed here in detail. The behavior of the magnetic submodes is alike. Therefore, the corresponding discussion of these modes will be rather brief.

The key quantities of the Mie solution are the properly normalized complex amplitudes of the scattered $a_n,\;b_n$ and internal $c_n,\;d_n$ field components (scattering coefficients). These coefficients should satisfy the boundary conditions, following from the continuity of the tangential components of the electric and magnetic fields at the surface of the particle. The conditions are split into two independent pairs for the electric and magnetic modes, respectively~\cite{Bohren::1998}
\begin{eqnarray}
  & & \xi_n(x)a_n + \psi_n(mx)d_n= \psi_n(x), \label{Eq:BC1electr} \\
  & & m\xi'_n(x)a_n +\psi'_n(mx)d_n = m \psi'_n(x), \label{Eq:BC2electr}\\
  & & m\xi_n(x)b_n + \psi_n(mx)c_n = m\psi_n(x), \label{Eq:BC1magn}\\
  & & \xi'_n(x)b_n + \psi'_n(mx)c_n = \psi'_n(x) \label{Eq:BC2magn}.
\end{eqnarray}
Here $x=2\pi R/\lambda$ is the size parameter; $\psi_n(z)$, \mbox{$\xi_n(z) = \psi_n(z) - i\chi_n(z)$}, $\psi_n(z)=zj_n(z)$ and $\chi_n(z)=-zy_n(z)$ are the Riccati-Bessel functions; $j_n(z),\;y_n(x)$ stand for the spherical Bessel functions~\cite{Bohren::1998}; ${}'=\partial/\partial z$ designates derivative with respect to the entire argument. Integer $n$ indicates the multipolarity of the mode, so that $n=1,\;2...$ correspond to the dipole mode, quadrupole mode, etc. It should be stressed that functions $\psi_n(z)$ and $\chi_n(z)$ are real for \mbox{real $z$.}

Solving Eqs.~\eqref{Eq:BC1electr}, \eqref{Eq:BC2electr} with respect to $a_n$, $d_n$, we obtain~\cite{Bohren::1998}
\begin{eqnarray}
  a_n &=& \frac{m\psi_n(mx)\psi_n'(x) - \psi_n(x)\psi_n'(mx)}{m\psi_n(mx)\xi_n'(x) - \xi_n(x)\psi_n'(mx)} \label{Eq:a_n} \\
  d_n &=& \frac{im}{m\psi_n(mx)\xi_n'(x) - \xi_n(x)\psi_n'(mx)},\label{Eq:d_n}
\end{eqnarray}
where the identity
\begin{equation}\label{Eq:identity_1}
  \psi_n(x)\xi'_n(x) - \psi'_n(x)\xi_n(x) \equiv i,
\end{equation}
following from the expression for the Wronskian of the Riccati-Bessel functions~\cite{Abramowitz::1965}, has been employed.

The corresponding treatment of the magnetic modes yields
\begin{eqnarray}
  b_n &=& \frac{m\psi_n(x)\psi'_n(mx)-\psi_n(mx)\psi'_n(x)}{m\xi_n(x)\psi'_n(mx)-\psi_n(mx)\xi'_n(x)} \label{Eq:b_n} \\
  c_n &=& -\frac{im}{m\xi_n(x)\psi'_n(mx)-\psi_n(mx)\xi'_n(x)} \label{Eq:c_n}
\end{eqnarray}

Note, that according to Eqs.~\eqref{Eq:d_n}, \eqref{Eq:identity_1}, \eqref{Eq:c_n} $d_n \equiv c_n \equiv 1$ at $m=1$, i.e., when the optical properties of the particle are identical to those of the embedding medium. Thus, $|d_n|,\;|c_n|$ may be regarded as the enhancement parameters for the field within the particle.

Eqs.~\eqref{Eq:a_n}, \eqref{Eq:d_n}, \eqref{Eq:b_n}, \eqref{Eq:c_n} may be written in the equivalent, identical form:
\begin{eqnarray}
a_n & = & \frac{F_n^{(a)}}{F_n^{(a)}+i G_n^{(a)}}\;,\;\;d_n=\frac{i\;m}{F_n^{(a)}+i G_n^{(a)}},\label{Eq:an_dn_FG}\\
b_n & = & \frac{F_n^{(b)}}{F_n^{(b)}+i G_n^{(b)}}\;,\;\;c_n=-\frac{i\;m}{F_n^{(b)}+i G_n^{(b)}},
\end{eqnarray}
with
\begin{eqnarray}
F_n^{(a)}&=&m\psi'_n(x)\psi_n(m x)-\psi_n(x)\psi'_n(m x)\;,\label{Eq:Fna}\\
G_n^{(a)}&=&-m\chi'_n(x)\psi_n(m x)+\chi_n(x)\psi'_n(m x)\;,\label{Eq:Gna}\\
F_n^{(b)}&=&m\psi_n(x)\psi'_n(mx)-\psi_n(mx)\psi'_n(x)\;,\label{Eq:Fnb}\\
G_n^{(b)}&=&-m\chi_n(x)\psi'_n(mx) + \chi'_n(x)\psi_n(mx). \label{Eq:Gnb}
\end{eqnarray}

Let us focus on the electric modes. At the limit of the large refractive index  we can keep just the first terms in Eqs.\eqref{Eq:Fna}--\eqref{Eq:Gna}:
\begin{eqnarray}
F_n^{(a)}\xrightarrow[{m \gg 1}]{}m\psi'_n(x)\psi_n(m x)\;,\label{Eq:Fna_m->infty}\\
G_n^{(a)}\xrightarrow[{m \gg 1}]{}-m\chi'_n(x)\psi_n(m x)\;. \label{Eq:Gna_m->infty}
\end{eqnarray}
Eqs. \eqref{Eq:Fna_m->infty}--\eqref{Eq:Gna_m->infty} lead to two important conclusions. First, the partial scattering coefficients $a_n$ converge to the $m$-{\it independent} form valid for the PRS:
\begin{equation}
a_n\xrightarrow[{m \gg 1}]{}a_n^{\rm (PRS)}\!=\!\frac{F_n^{(a,{\rm PRS})}}{F_n^{(a,{\rm PRS})}+i G_n^{(a,{\rm PRS})}}\!=\!\frac{\psi'_n(x)}{\xi'_n(x)},
\label{Eq:a_n-PRS}
\end{equation}
with
\begin{equation}\label{Eq:Fna_PRS-Gna_PRS}
  F_n^{(a,{\rm PRS})}=\psi'_n(x),\;\;G_n^{(a,{\rm PRS})}=-\chi'_n(x).
\end{equation}
Note that expression, Eq.~\eqref{Eq:a_n-PRS} can be also obtained from the boundary condition, Eq.~\eqref{Eq:BC2electr}, if we suppose there $d_n = 0$, as it should be for a PRS, or by direct solution of the light scattering problem for a PRS~\footnote{The problem of light scattering by a PRS is a bit tricky. The point is that the field inside the PRS is zero identically. Then, instead of the four independent scattering coefficients: two for the scattered field outside the particle and two for the field within the particle, only the former two remain. The reduction of the independent constants of integration of the Maxwell equations, requires the corresponding reduction of the boundary conditions. It could be shown, that for the electric modes discussed here only boundary condition, Eq.~\eqref{Eq:BC2electr} remains, while the boundary condition, Eq.~\eqref{Eq:BC1electr} should be removed at this limit.}.

Second, the internal coefficients $d_n$ in this limit still keep the dependence on $m$, converging to
\begin{eqnarray}\label{Eq:d_n_at_m->infty}
d_n\xrightarrow[m \gg 1]{}d_n^{(\lim)}=\frac{i}{\psi_n(mx)\xi'_n(x)}.
\end{eqnarray}
\noindent
 It is relevant to mention that while expressions, Eqs.~\eqref{Eq:a_n-PRS}, \eqref{Eq:d_n_at_m->infty} satisfy boundary condition, Eq.~\eqref{Eq:BC1electr} identically, boundary condition, Eq.~\eqref{Eq:BC2electr} for these expressions is satisfied only asymptotically at \mbox{$m \rightarrow \infty$.}

It is interesting to note also, that the limit, Eqs.~\eqref{Eq:a_n-PRS}, \eqref{Eq:Fna_PRS-Gna_PRS} corresponds to the scattering coefficient for a sphere made of a perfect electric conductor ~\cite{Balanis:Book:2012}. However, in the problem in question a sphere made of \emph{a perfect insulator} with zero conductivity (Im$\,\varepsilon = 0$) is considered. The coincidence of the two limits occurs owing to the fact, that despite the conductivity current in our case is zero, the displacement current plays its role. If the field {\bf E} inside the particle does not vanish, the displacement current diverges at Re$\,\varepsilon \rightarrow \infty$. It brings about exactly the same result as that, following from the divergence of the conductivity current in a perfect electric conductor.

It should be stressed that limit, Eq.~(\ref{Eq:a_n-PRS}) is not valid in the vicinities of the points, where $\psi_n(mx)=0$. Moreover, exactly at these points the internal coefficient diverges $d_n^{(\lim)}\rightarrow\infty$, see Eq.~(\ref{Eq:d_n_at_m->infty}). It allows us to conclude that the condition
\begin{equation}\label{Eq:psi=0}
  \psi_n(mx)=0
\end{equation}
corresponds to the resonances excited in a HRI dielectric sphere at large enough $m$. This resonance condition can be further simplified taking into account that in the range $m x>n^2$ (known as the Fraunhofer regime) the Riccati-Bessel functions are reduced to simple trigonometrical ones, so that the condition $\psi_n(mx)=0$ reads
\begin{eqnarray}
\psi_n(m x)\cong\sin\left(m x-\frac{n\pi}{2}\right)=0 \label{Eq:sin=0}\;.
\end{eqnarray}
An analogous treatment of the magnetic coefficients $b_n$ (for the outer problem) and $c_n$ (for the inner) brings about the following expressions:
\begin{eqnarray}
F_n^{(b)}\xrightarrow[{m \gg 1}]{}m\psi_n(x)\psi'_n(m x)\;,\label{Eq:Fnb_m->infty}\\
G_n^{(b)}\xrightarrow[{m \gg 1}]{}-m\chi_n(x)\psi'_n(m x)\;. \label{Eq:Gnb_m->infty}
\end{eqnarray}
\begin{equation}
b_n\xrightarrow[{m \gg 1}]{}b_n^{\rm (PRS)}\!=\!\frac{F_n^{(b,{\rm PRS})}}{F_n^{(b,{\rm PRS})}+i G_n^{(b,{\rm PRS})}}\!=\!\frac{\psi_n(x)}{\xi_n(x)},
\label{Eq:b_n-PRS}
\end{equation}
\begin{equation}\label{Eq:Fnb_PRS-Gnb_PRS}
  F_n^{(b,{\rm PRS})}=\psi_n(x),\;\;G_n^{(b,{\rm PRS})}=-\chi_n(x),
\end{equation}
\begin{eqnarray}\label{Eq:c_n_at_m->infty}
c_n\xrightarrow[m \gg 1]{}c_n^{(\lim)}=-\frac{i}{\psi'_n(mx)\xi_n(x)},
\end{eqnarray}
\noindent
with the following condition for the magnetic modes resonances:
\begin{eqnarray}
\psi'_n(m x)=0\;, \label{Eq:psi_prime=0}
\end{eqnarray}
leading in the Fraunhofer regime to the equation
\begin{equation}\label{Eq:cos=0}
 \psi'_n(m x)\cong \cos\left(m x-\frac{n\pi}{2}\right)=0.
\end{equation}

The two resonance conditions may be rewritten in a unified form:
\begin{eqnarray}
& & m^{({\rm res},E)}_{n,p}\cong(n+2p)\frac{\pi}{2x}, \label{Eq:m_res_E}\\
& & m^{({\rm res},H)}_{n,p}\cong(n+2p+1)\frac{\pi}{2x} \label{Eq:m_res_H}.
\end{eqnarray}
where $p$ is a non-negative integer number. The meaning of resonance conditions, Eqs.~\eqref{Eq:m_res_E}, \eqref{Eq:m_res_H} becomes absolutely clear, if we recall that $x = 2\pi R/\lambda$ and $\lambda/m$ is the wavelength inside the scattering particle. Then, Eqs.~\eqref{Eq:m_res_E}, \eqref{Eq:m_res_H} may be rewritten as follows:
\begin{eqnarray*}
  & & \frac{\lambda}{m^{({\rm res},E)}_{n,p}}\left(\frac{n}{2}+p\right) \cong 2R, \\
  & & \frac{\lambda}{m^{({\rm res},H)}_{n,p}}\left(\frac{n+1}{2}+p\right) \cong 2R,
\end{eqnarray*}
i.e., the resonances occur when an integer number of the half-waves equals the diameter of the sphere, cf.~\cite{Kuznetsov:SR:2012}.

These resonance conditions lead to a number of interesting conclusions, cf. \cite{Hulst::1981}. First, each multipole has, in general, infinite number of resonances, associated with $p=0,1,2,\ldots$. Second, owing to the additional degree of freedom related to variations of $p$, there is the \emph{multiple degeneracy} of the resonances. Specifically, the resonances with different multipolarity $n$ occur at one and the same value of $m$, provided for these resonances the variation of $n$ is compensated by the corresponding variation of $p$, i.e., $n+2p$ for the electric resonant modes remains the same and equals $n+1+2p$ for magnetic.

Next, at a give $n$ the points of the $n$th electric resonances correspond to those of the $(n+1)th$ magnetic and vice versa. Note also that at a fixed $n$ the points of resonances of one type (i.e., either electric, or magnetic) are situated in the $m$-axis just in the middle of the spacing between the points of the other type, so that the maxima for the electric modes correspond to the minima of magnetic and the other way around, see Eqs.~\eqref{Eq:m_res_E}, \eqref{Eq:m_res_H}.

\begin{figure}
  \includegraphics[width=\columnwidth]{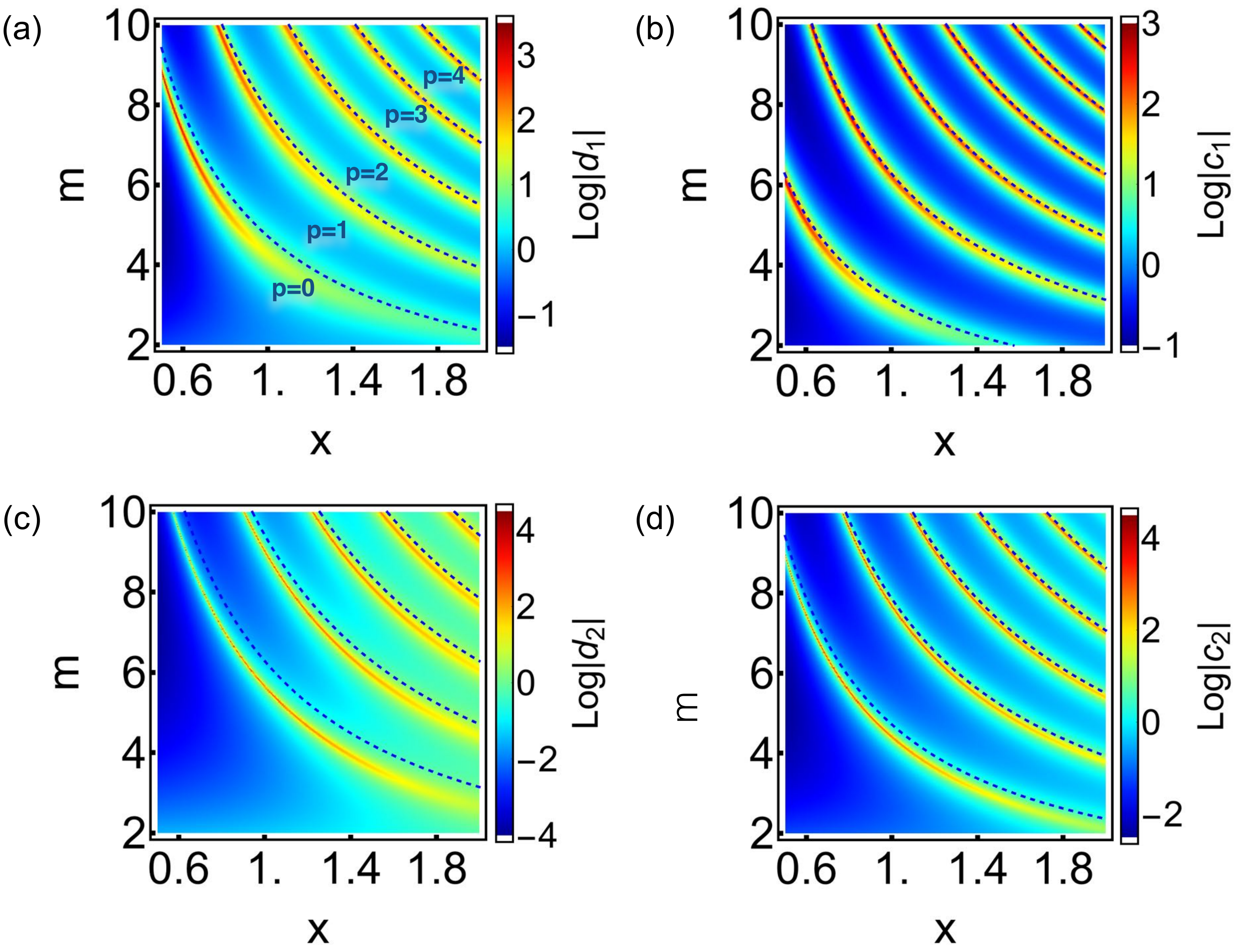}\\
  \caption{(color online) Dependence of electric $d_{1,2}$ (a,c) and magnetic $c_{1,2}$ (b,d) internal coefficients vs size parameter $x$ and refractive index $m$, calculated according to the exact Mie solution. Dashed lines indicate the asymptotic resonance conditions, obtained in the limit $m\rightarrow\infty$ , see Eqs.~(\ref{Eq:m_res_E}), (\ref{Eq:m_res_H}) at $p = 0,\,1,\,2,\,3.$ Despite the employed values of $m$ are not so large, the agreement between the dashed lines and the positions of the actual resonances is very good.
}\label{fig:fig2}
\end{figure}

In Fig.~\ref{fig:fig2} we plot the dependence of the first two electric $d_{1,2}$ and magnetic $c_{1,2}$ internal coefficients vs. size parameter $x$ and refractive index $m$, calculated according to the exact Mie solution. In this figure we also plot asymptotic resonance conditions, Eqs.~(\ref{Eq:m_res_E}), (\ref{Eq:m_res_H}). The agreement between the position of the resonances according to the exact solution and approximate conditions, Eqs.~(\ref{Eq:m_res_E}), (\ref{Eq:m_res_H}) is surprisingly good.

It should be stressed, however, that the results obtained do not mean a strong overlap of the resonances yet. The point is that the right-hand-sides of Eqs.~\eqref{Eq:psi=0}, \eqref{Eq:psi_prime=0} are just the first terms of the corresponding asymptotic expansions in powers of small $1/(mx)$. Higher order terms, dropped in these equations bring about the mismatches between the points of the resonances for different modes. To claim the strong overlap of the resonances we must make sure that the mismatches are smaller than the corresponding linewidths. Let us proceed to a discussion of these issues.

\section{Linewidth and lineshape}\label{sec:3}

To understand the meaning of the existence of the limit at $m \rightarrow \infty$ for scattering coefficients $a_n$, and its absence for internal ones, $d_n$, we have to inspect Eqs.~\eqref{Eq:Fna}--\eqref{Eq:Gna} more carefully.

First of all note, direct calculations of $a_n$ based upon Eqs. \eqref{Eq:an_dn_FG}, \eqref{Eq:Fna_m->infty}, \eqref{Eq:Gna_m->infty}, \eqref{Eq:psi=0}, \eqref{Eq:m_res_E} brings about the uncertainty of the type 0/0. To resolve the lineshape, let us consider a small departure of the refractive index from a certain resonant value: $\delta m = m - m^{\rm (res)}$, where $m^{\rm (res)}$ satisfies Eq.~\eqref{Eq:psi=0}.  Here and in what follows for simplicity of notations we put $m^{\rm (res)}$ for $m^{({\rm res}, E)}_{n,p}$. Then, bearing in mind that in the vicinity of the resonance $\psi_n\left((m^{{\rm (res)}}+\delta m)x\right) \cong \psi'_n(m^{{\rm (res)}}x)x\delta m$,
we readily obtain the following formulas for $F^{(a)}_n\!\!,\; G^{(a)}_n$:
\begin{eqnarray}
  \!\!\!\!\!\!\!\!F^{(a)}_n &\cong& \psi_n'(m^{\rm (res)}x)\left\{m^{\rm (res)}\psi_n'(x)x\delta m - \psi_n(x) \right\}\!, \label{Eq:Fn_res_a} \\
  \!\!\!\!\!\!\!\!G^{(a)}_n &\cong& -\psi_n'(m^{\rm (res)}x)\left\{m^{\rm (res)}\chi_n'(x)x\delta m - \chi_n(x) \right\}\!. \label{Eq:Gn_res_a}
\end{eqnarray}
It is seen from Eqs. \eqref{Eq:an_dn_FG} \eqref{Eq:Fn_res_a}, \eqref{Eq:Gn_res_a}, that scattering coefficient reaches its maximum (the points of the constructive interference), $a_n =1$ at
\begin{equation}\label{Eq:delta_m_G_a}
\delta m^{(a)}_G \cong \frac{1}{x m^{\rm (res)}}\frac{\chi_n(x)}{\chi_n'(x)},
\end{equation}
and minimal value, $a_n=0$ (the points of the destructive interference) at
\begin{equation}\label{Eq:delta_m_F_a}
\delta m^{(a)}_F \cong \frac{1}{x m^{\rm (res)}}\frac{\psi_n(x)}{\psi_n'(x)}.
\end{equation}

The corresponding expressions for $b_n$ read
\begin{eqnarray}
  F^{(b)}_n &\cong& m^{\rm (res)}\psi_n(x)\psi''_n(m^{\rm (res)}x)x\delta m  \label{Eq:Fn_res_b}\\
   & & - \psi'_n(x)\psi_n(m^{\rm (res)}x),  \nonumber\\
  G^{(b)}_n &\cong& -m^{\rm (res)}\chi_n(x)\psi''_n(m^{\rm (res)}x)x\delta m \label{Eq:Gn_res_b}\\
   & & + \chi'_n(x)\psi_n(m^{\rm (res)}x) \nonumber.
\end{eqnarray}
\begin{equation}\label{Eq:delta_m_G_b}
\delta m^{(b)}_G \cong \frac{1}{x m^{\rm (res)}}\frac{\chi'_n(x)\psi_n(m^{\rm (res)}x)}{\chi_n(x)\psi''_n(m^{\rm (res)}x)},
\end{equation}
\begin{equation}\label{Eq:delta_m_F_b}
\delta m^{(b)}_F \cong \frac{1}{x m^{\rm (res)}}\frac{\psi'_n(x)\psi_n(m^{\rm (res)}x)}{\psi_n(x)\psi''_n(m^{\rm (res)}x)}.
\end{equation}
where now $m^{{\rm (res)}}$ stands for $m^{({\rm res},H)}_{n,p}\!\!\!,$ satisfying Eq.~\eqref{Eq:psi_prime=0}. Note, $\psi_n(m^{\rm (res)}x)/\psi''_n(m^{\rm (res)}x) \cong -1$ at $m^{\rm (res)}x > n^2$, see Eqs.~\eqref{Eq:sin=0}.

\begin{figure}
\centering
\includegraphics[width=\linewidth]{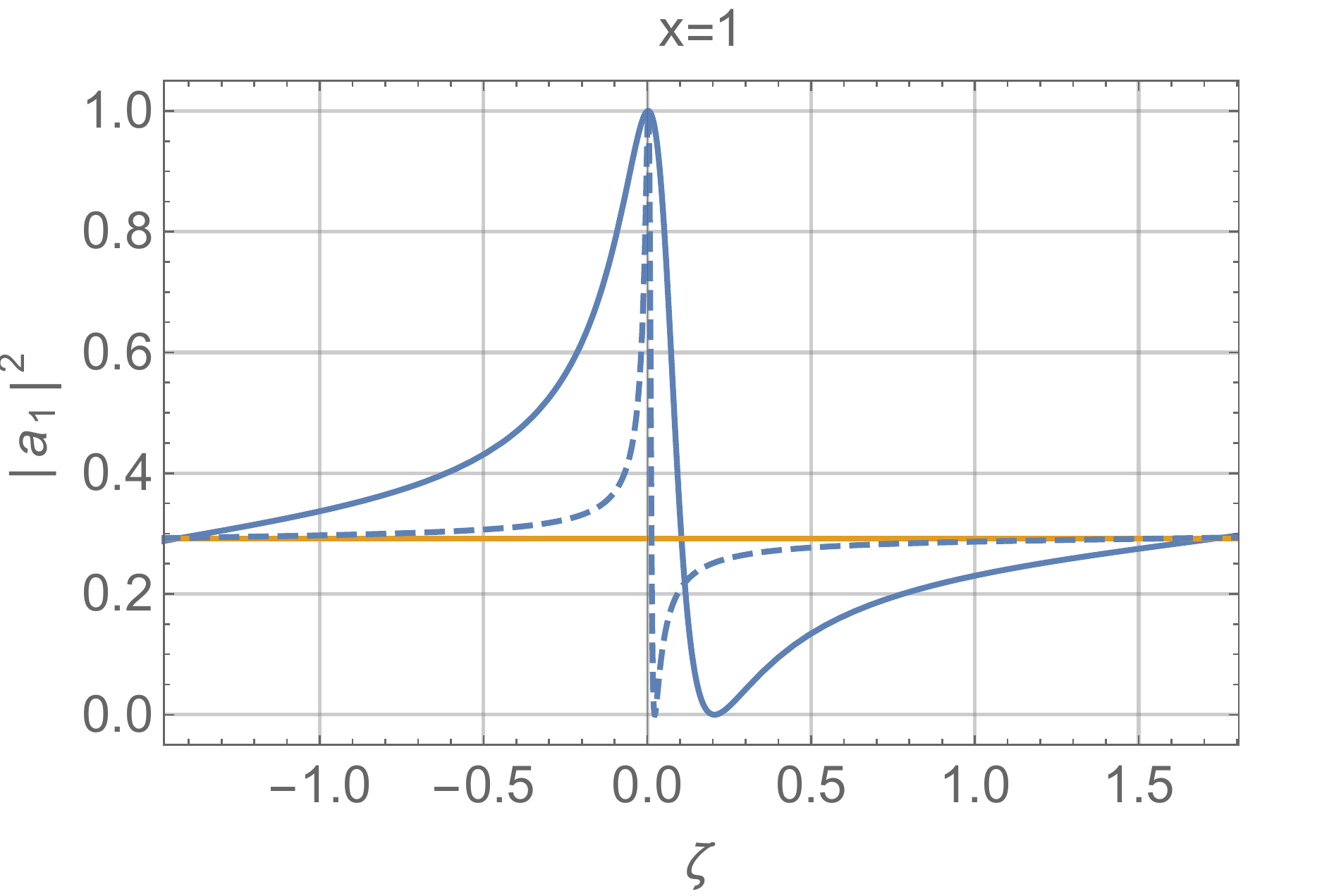}
\caption{(color online) Contraction of the Fano resonance lines for $|a_1|^2$ with an increase in $m;\;x=1;\;\zeta = m - m^{{\rm (res)}}.$ Here \mbox{$m^{{\rm (res)}}= 10.75$} (solid line) and 111.5 (dashed line).  A yellow horizontal line corresponds to $a_1^{\rm (PRS)}$. }
\label{fig:a1}
\end{figure}

The obtained results explain the mentioned convergence of $a_n$ to the $m$-independent form at $m \rightarrow \infty$: at an increase in $m$ the width of the resonance line $|\delta m_G - \delta m_F|$ contracts as $1/xm^{\rm (res)}$, while the amplitude of the resonances and spacing between two adjacent maxima (minima), both remain $m$-independent. Asymptotically, at $m \rightarrow \infty$ the maxima and minima merge with each other, and the resonance profile vanishes. An example of such a process for $|a_1|^2$ at $x=1$ is presented in Fig.~\ref{fig:a1}.

Now let us inspect coefficients $d_n$. As it follows from \mbox{Eqs.~\eqref{Eq:BC2electr}, \eqref{Eq:a_n-PRS}}
\begin{equation}\label{Eq:a_n_trough_aPRS_and_d_n}
  a_n = a_n^{\rm (PRS)} -\frac{\psi_n'(mx)}{m\xi_n'(x)}d_n.
\end{equation}
Then, it is convenient to introduce the rescaled coefficient
\begin{equation}\label{Eq:d_n_tilde}
  \tilde{d_n} = \frac{\psi_n'(mx)}{m\xi_n'(x)}d_n \equiv \frac{i}{m\frac{\psi_n(mx)}{\psi_n'(mx)} - \frac{\xi_n(x)}{\xi_n'(x)}}\frac{1}{\xi'^2_n(x)}.
\end{equation}
Next, utilizing the identity
\begin{equation}\label{Eq:identity_2}
  \left|\xi'^2_n(x){\rm Im}\,\left(\frac{\xi_n(x)}{\xi_n'(x)}\right)\right| \equiv 1,
\end{equation}
see Appendix A, its modulus may be presented as follows:
\begin{equation}\label{Eq:d_n_tilde_AB}
  |\tilde{d_n}|^2 = \frac{1}{\left(m\frac{\psi_n(mx)}{\psi'_n(mx)}-A_n^{(d)}(x)\right)^2B_n^{(d)2}(x)+1},
\end{equation}
where
\begin{equation}\label{Eq:A_n^{(d)}_B_n^{(d)}}
  A_n^{(d)}(x) = {\rm Re}\,\left(\frac{\xi_n(x)}{\xi_n(x)'}\right);\;\;\; B_n^{(d)}(x) = \left|\xi_n'^2(x)\right| \equiv \left|\xi_n'(x)\right|^2.
\end{equation}

The conclusion, which immediately follows from Eq.~\eqref{Eq:A_n^{(d)}_B_n^{(d)}}, is that the resonant values of $|\tilde{d_n}|^2$ equals unity and the resonances are achieved at the values of $m$, satisfying the equation:
\begin{equation}\label{Eq:resonant_cond_|d_n|}
  m\frac{\psi_n(mx)}{\psi'_n(mx)}=A_n^{(d)}(x).
\end{equation}

We should emphasize that the resonance condition stipulated by Eq.~\eqref{Eq:resonant_cond_|d_n|} is an \emph{exact result}, valid at any $m$ and $x$. To obtain the lineshape in the proximity of $m=m^{\rm (res)}$, with $m^{\rm (res)}$ defined by Eq.~\eqref{Eq:psi=0}, as usual, we have to expand the argument of $\psi_n(mx)$ in powers of a small $\delta m = m-m^{\rm (res)}$, so that $\psi_n(mx) \cong \psi'_n(m^{\rm (res)}x)x\delta m$. Then, the general expression, Eq.~\eqref{Eq:d_n_tilde_AB} is reduced to the following simple form:
\begin{equation}\label{Eq:lineshape_|d_n|^2}
  |\tilde{d_n}|^2 = \frac{1}{\left(m^{\rm (res)}x\delta m -A_n^{(d)}(x)\right)^2B_n^{(d)2}(x)+1}.
\end{equation}
It is a typical Lorentzian profile with a maximum at
\begin{equation}\label{Eq:delta_m_|d|_max}
  \delta m_{|d_n|^2}^{\rm (res)} = \frac{A_n^{(d)}(x)}{m^{\rm (res)}x}
\end{equation}
and the half-maximum linewidth (FWHM)
\begin{equation}\label{Eq:linewidth_|d_n|^2}
  \gamma_{|d_n|^2} = \frac{2}{m^{({\rm res})}xB_n^{(d)}(x)}.
\end{equation}
Note, that Eq.~\eqref{Eq:delta_m_|d|_max} gives the mismatch between the positions of the resonance points defined according to Eq.~\eqref{Eq:psi=0} and the ones corresponding to the maxima of $|d_n|^2$.

It seems we have encountered a paradox. On one side, we have obtained that in contrast to the outer problem, the inner one does not have a definite limit at $m \rightarrow \infty$, see Eq.~\eqref{Eq:d_n_at_m->infty}.

On the other side, $\tilde{d_n} = a_n^{\rm (PRS)} - a_n$, see Eqs.~\eqref{Eq:a_n_trough_aPRS_and_d_n},~\eqref{Eq:d_n_tilde}. This equality means that since $a_n \xrightarrow[m\rightarrow\infty]{}a_n^{\rm (PRS)}$, coefficient $\tilde{d_n}$ should vanish in this limit. The latter reasoning agrees well with the just obtained for $\tilde{d_n}$ lineshape. Indeed, being equal in its maximum to unity, the linewidth for $|\tilde{d_n}|^2$ vanishes at $m \rightarrow \infty$ as $1/m^{\rm (res)}$, while the value of $|\tilde{d_n}|^2$ at off-resonance regions vanishes as $1/m^2$, see Eq.~\eqref{Eq:d_n_tilde_AB}.

Then, at $m \rightarrow \infty$ the resonance lines become infinitesimally narrow and the entire profile $|\tilde{d_n}(m)|^2$ vanishes. Finally, because the difference between $\tilde{d_n}$ and ${d_n}$ is just in the multiplicative scaling factor the same conclusion, seemingly, may be applied to the profile $|{d_n(m)}|^2$.

However, one should be careful here, because the scaling factor itself depends on $m$. Since \mbox{$\max{|\tilde{d_n}|^2} = 1$}, the maximal value of $|d_n|^2$ increases as $m^{(\rm{res})\,2}$, see Eq.~\eqref{Eq:d_n_tilde}. It means that the total area under a given resonance line increases with an increase in $m^{\rm (res)}$ linearly in $m^{\rm (res)}$, see Eq.~\eqref{Eq:linewidth_|d_n|^2}, making (in contrast to $a_n$) the resonances more and more pronounced.

Note, however, that it does not mean nonexistence of \emph{any} universality in the profile $|d_n(m)|^2$ at $m \rightarrow \infty$. Let us consider \emph{a part} of the profile from its bottom to any \emph{fixed} value $D_n^2$. According to Eq.~\eqref{Eq:d_n_tilde}
\begin{equation}\label{Eq:d_n_through_d_n_tilde}
  d_n =\frac{m\xi_n'(x)}{\psi_n'(mx)}\tilde{d}_n \equiv \frac{m(\psi_n'(x) -i\chi_n'(x))}{\psi_n'(mx)}\tilde{d}_n .
\end{equation}
Then, bearing in mind that in the Fraunhofer regime for the discussed resonances $|\psi_n'(m^{({\rm res})}x)| = 1$, see Eqs.~\eqref{Eq:sin=0}, \eqref{Eq:m_res_E}, the width of the line at this distance from the bottom ($\gamma_D$) is given by the difference \mbox{$|\delta m_1 - \delta m_2|$}, where $\delta m_{1,2}$ are the two roots of the equation
\begin{equation}\label{Eq:eq_for Gamma_D}
  \frac{m^{(\rm{res})\,2}\left(\psi_n'^2(x)+\chi_n'^2(x)\right)}{\left(m^{\rm (res)}x\delta m -A_n^{(d)}(x)\right)^2B_n^{(d)2}(x)+1} = D^2,
\end{equation}
\begin{figure}
\centering
\includegraphics[width=\linewidth]{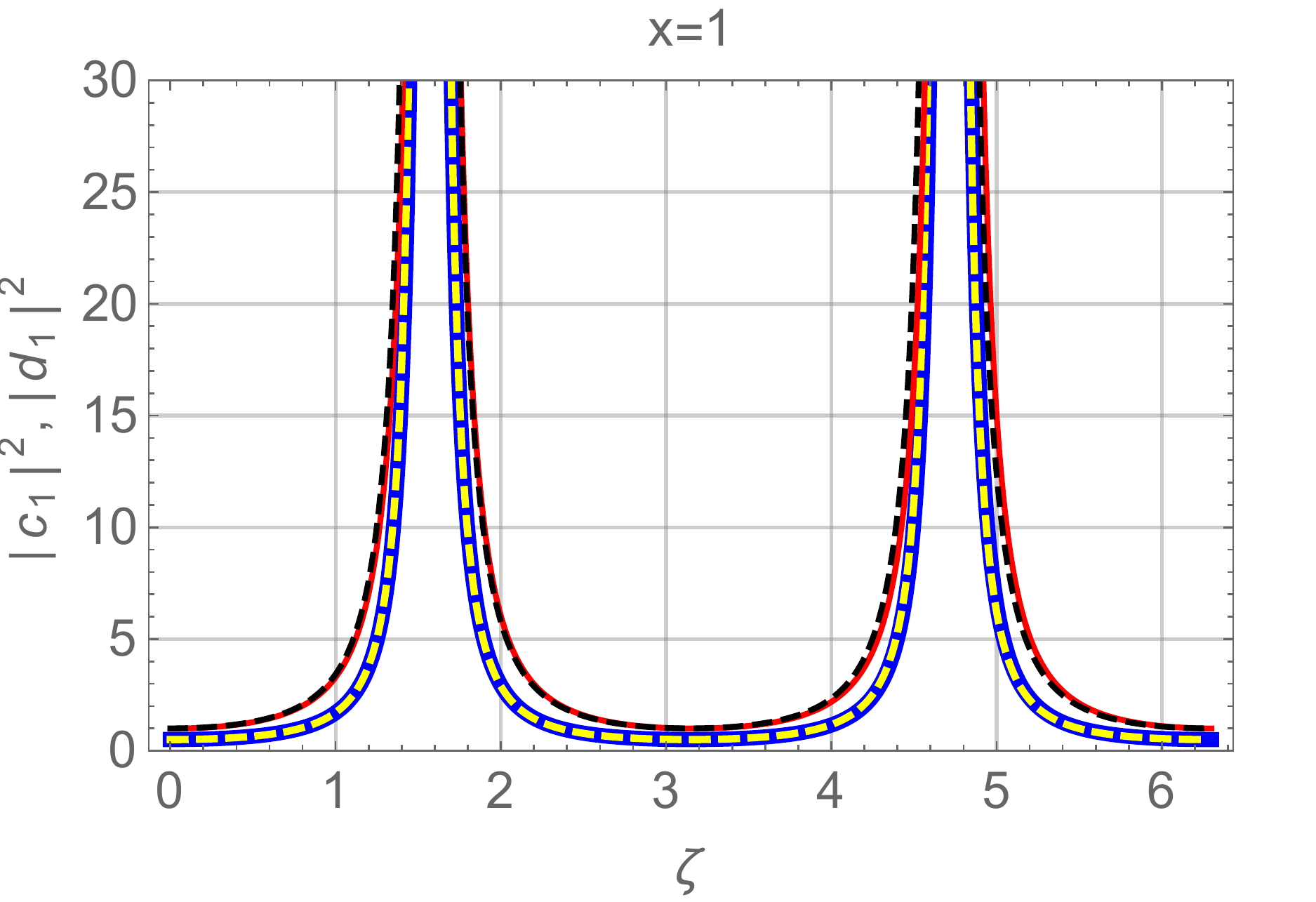}
\caption{(color online) Independence of $m^{{\rm (res)}}$ of the bottom parts of the resonance lines for profiles $|c_1|^2$ and $|d_1|^2$ at $x =1$ and large values of $m^{{\rm (res)}};\; \zeta = m - m^{{\rm (res)}}$. Here profiles $|c_1(\zeta)|^2$ and $|d_1(\zeta)|^2$ are presented at $m^{{\rm (res)}} = 14.1019...$ (thick blue solid line), $m^{{\rm (res)}} = 1013.1631...$ (thick yellow dashed line) and $m^{{\rm (res)}} = 15.5792$ (red solid line), $m^{{\rm (res)}} = 1014.7325...$ (black dashed line), respectively. Increase in $m^{{\rm (res)}}$ in the three orders of magnitudes, practically, does not affect the shape of the lines. Note the perfect coincidence of the corresponding lines, even in the off-resonance regions, despite Eq.~\eqref{Eq:Gamma_D} is valid only in the vicinity of the points of the resonances and, generally speaking, cannot be applied to describe the lineshape far from them.}
\label{fig:d1}
\end{figure}

Trivial calculations show that at \mbox{$D^2 \ll m^{(\rm{res})\,2}\left(\psi_n'^2(x)+\chi_n'^2(x)\right)$} the quantity $\gamma_D$ converges to the following $m$-independent expression:
\begin{equation}\label{Eq:Gamma_D}
  \gamma_D \cong \frac{2\sqrt{\psi_n'^2(x) + \chi_n'^2(x)}}{DB(x)x}.
\end{equation}
This asymptotics is valid for any $D$ satisfying the aforementioned constraint. It means that at large $m$ a part of the resonance lineshape close to the bottom of the resonance line becomes universal. An example of this universality for $n=1$ and $x=1$ is shown in Fig.~\ref{fig:d1}.

Thus, instead of vanishing {(as it should be for a PRS)} at $m \rightarrow \infty$ the resonance lines for coefficient $|d_n|^2$ at the bottom converge to a certain universal form, while the maximal value of $|d_n|^2$ at the peak of the resonance increases as $m^{(\rm{res})\,2}$ and the half-maximum linewidth contracts as $1/m^{\rm (res)}$. The behavior of $c_n$ is quite analogous to that of $d_n$. The expressions describing this behavior are presented below for references:
\begin{eqnarray}
 & & b_n = b_n^{\rm (PRS)} -\tilde{c}_n,\;\;\tilde{c}_n \equiv \frac{\psi_n(mx)}{m\xi_n(x)}c_n, \label{Eq:b_n_trough_bPRS_and_c_n--c_n^tilde} \\
 & & |\tilde{c_n}|^2  =  \frac{1}{\left(m\frac{\psi'_n(mx)}{\psi_n(mx)}-A_n^{(c)}(x)\right)^2B_n^{(c)2}(x)+1} \label{Eq:c_n_tilde_AB} \\
 & & \approx \frac{1}{\left(\frac{\psi''_n(m^{\rm (res)}x)}{\psi_n(m^{\rm (res)}x)}m^{\rm (res)}x\delta m -A_n^{(c)}(x)\right)^2B_n^{(c)2}(x)+1},\nonumber \\
 & & \delta m_{|c_n|^2}^{\rm (res)} = \frac{A_n^{(c)}(x)\psi_n(m^{\rm (res)}x)}{m^{\rm (res)}x\psi''_n(m^{\rm (res)}x)},\;
  \label{Eq:delta_m_|c|_max}\\
  & & \gamma_{|c_n|^2} = \frac{2}{m^{({\rm res})}xB_n^{(c)}(x)}\left|\frac{\psi_n(m^{\rm (res)}x)}{\psi''_n(m^{\rm (res)}x)}\right|. \label{Eq:linewidth_|c_n|^2}
\end{eqnarray}
Here $m^{\rm (res)}$ is defined according to Eq.~\eqref{Eq:psi_prime=0},
\begin{equation}\label{Eq:A_n^{(c)}_B_n^{(c)}}
  A_n^{(c)}(x) = {\rm Re}\,\left(\frac{\xi'_n(x)}{\xi_n(x)}\right);\;\;\; B_n^{(c)}(x) = \left|\xi_n^2(x)\right| \equiv \left|\xi_n(x)\right|^2,
\end{equation}
and to obtain Eq.~\eqref{Eq:c_n_tilde_AB} the identity
\begin{equation*}
   \left|\xi^2_n(x){\rm Im}\,\left(\frac{\xi'_n(x)}{\xi_n(x)}\right)\right| \equiv 1,
\end{equation*}
whose proof is completely analogous to that for Eq.~\eqref{Eq:identity_2}, has been employed.

Let us make numerical estimates. At \mbox{$x=1.3$} and \mbox{$m \approx 3\sim5$} (typical values for a number of semiconductors in the visible and IR diapason) the resonant \mbox{$|d_1|^{\rm (res)} \approx 3.01$} \mbox{$(m^{\rm (res)} \approx 3.33)$};  \mbox{$|d_2|^{\rm (res)} \approx 11.78$} \mbox{$(m^{\rm (res)} \approx 4.27)$}; \mbox{$|d_3|^{\rm (res)} \approx 93.59$}; \mbox{$(m^{\rm (res)} \approx 5.30)$}. It gives an estimate for the range of the growth of the electric field inside the particle with respect to the incident wave, while the corresponding values for the volume density of the electromagnetic energy will be of the orders of squires of these values.

To conclude this section a certain important remark should be made. Though it is convenient for the theoretical study adopted in the present paper to inspect the resonances at varying $m$ and fixed values of the other problem parameters, such an approach seems completely irrelevant from the physical viewpoint. Really, while in an actual experiment, it is rather easy to change the value of the size parameter, $x$ just varying the wavelength of the incident light, it is very difficult to change $m$. The refractive index is a given property of the material the particle made of. To change it either materials with strong dispersion should be employed, or one has to have a set of particles with the same size but different refractive indices varying in small steps. Both the options look unrealistic.

Thus, it seems our study is quite meaningless. Fortunately, this is not the case. The point is that the actual parameter of the theory is the product, $\rho = mx$. The problem in question corresponds to $\rho \gg 1$, while the spacing between two sequential resonances $\delta \rho = \pi$, see Eqs.~\eqref{Eq:m_res_E}, \eqref{Eq:m_res_H}. Now, if instead of variations of $m$ at a fixed $x$ we consider variations of $x$ at a fixed $m$, to cover the distance between the two sequential resonances we have to consider departures of $x$ from a resonant value $x^{{\rm(res)}}$ (defined by the same conditions, Eqs.~\eqref{Eq:psi=0}, \eqref{Eq:psi_prime=0}) of the order $\delta x \sim 1/m \ll 1$. It means that in slowly-varying functions of $x$ solely, namely in $\xi_n(x),\;\psi_n(x)$ and $\chi_n(x)$ we may neglect these small variations of $x$, replacing these functions by their values at $x = x^{{\rm(res)}}$. The only remaining step is to replace $\delta \rho = x\delta m$ by $\delta \rho = m\delta x$. Then, all the expressions obtained in this section and in what follows are readily recalculated for the case of varying $x$ and fixed $m$.

\section{ Fano resonances in partial wave scattering}

Up to now we have inspected the modula of the scattering coefficients. Let us focus on their phases. Following the approach, described in monograph~\cite{Hulst::1981}, it is convenient to introduce a real angle $\Delta_n^{(a)}$ according to the expression
\begin{eqnarray}\label{Eq:Delta}
\tan\Delta_n^{(a)} \equiv \frac{F_n^{(a)}}{G_n^{(a)}}\;.
\end{eqnarray}
\mbox{}\\

Then, $a_n$ can be written in the following form
\begin{eqnarray}\label{Eq:Sca2}
a_n=\frac{\tan\Delta_n^{(a)}}{\tan\Delta_n^{(a)}+i}=\frac{i}{2}(1-e^{2i\Delta_n^{(a)}})\;.
\end{eqnarray}
Let us recall now that $a_n$ may be presented as a sum of two terms: $a_n^{\rm (PRS)}$ and $\tilde{d}_n=d_n\psi_n'(mx)/(m\xi_n'(x))$, see Eq.~\eqref{Eq:a_n_trough_aPRS_and_d_n}. Here the first term does not depend on $m$, while the second is a sharp function of $m$. Then, it makes sense to split $\Delta_n^{(a)}$ into two corresponding parts, namely to present it as
\begin{equation}\label{Eq:Delta_as_a_sum}
 \Delta_n^{(a)} \equiv \Delta_n^{(a,{\rm PRS})} + \Delta_n^{(a,{\rm res})},
\end{equation}
where $\Delta_n^{(a,{\rm PRS})}$ is defined for a PRS in the same manner as that for the just introduced $\Delta_n^{(a)}$ for a particle with arbitrary $m$, i.e.,
\begin{equation}\label{Eq:tan_delta_PRS}
  \tan \Delta_n^{(a,{\rm PRS})} \equiv \frac{F_n^{(a,{\rm PRS})}}{G_n^{(a,{\rm PRS})}} = - \frac{\psi_n'(x)}{\chi_n'(x)},
\end{equation}
see Eq.~\eqref{Eq:Fna_PRS-Gna_PRS}. Then, the quantity $\Delta_n^{(a,{\rm res)}}$ is defined by the identity
\begin{eqnarray}
\tan\Delta_n^{(a)} &\equiv& \tan(\Delta_n^{(a,{\rm res)}} + \Delta_n^{(a,{\rm PRS})})=\nonumber \\
& & \frac{\tan\Delta_n^{(a,{\rm res)}} +\tan\Delta_n^{(a,{\rm PRS})}}{1-\tan\Delta_n^{(a,{\rm res)}}\tan\Delta_n^{(a,{\rm PRS)}}}\;.\label{Eq:Delta_res}
\end{eqnarray}

Taking into account Eqs.~\eqref{Eq:identity_1}, \eqref{Eq:an_dn_FG} and \eqref{Eq:tan_delta_PRS}, after some algebra it is possible to show that to satisfy Eq.~\eqref{Eq:Delta_res} identically, tangent of $\Delta_n^{(a,{\rm res)}}$ must be equal
\begin{equation}\label{Eq:tan_Delta_res}
  \tan \Delta_n^{(a,{\rm res)}} =-\frac{\psi'_n(mx)}{G^{(a,{\rm PRS})}_n G_n^{(a)}+F^{(a,{\rm PRS})}_n F_n^{(a)}}.
\end{equation}

If now we introduce the notations
\begin{equation}\label{Eq:q_and_epsilon_a}
  q_n^{(a)} \equiv -\cot \Delta_n^{(a,{\rm PRS})}= \frac{\chi_n'(x)}{\psi_n'(x)},\;\; \epsilon_n^{(a)} \equiv -\cot \Delta_n^{(a,{\rm res)}},
\end{equation}
the expression for $|a_n|^2$ transforms into the following:

\begin{eqnarray}
|a_n|^2=\frac{\left(\epsilon_n^{(a)}+q_n^{(a)}\right)^2}{\left(1+q^{(a)2}\right)\left(1+\epsilon^{(a)2}\right)}\;.\label{Eq:|a_n|_Fano}
\end{eqnarray}
It is the conventional Fano profile, normalized to its maximal value,~\cite{Fano:PR:1961}.

A similar treatment of $b_n$ brings about analogous results with
\begin{equation}\label{Eq:qb_n}
  q_n^{(b)} = \frac{\chi_n(x)}{\psi_n(x)}.
\end{equation}
\begin{figure}
  \centering
  \includegraphics[width=\linewidth]{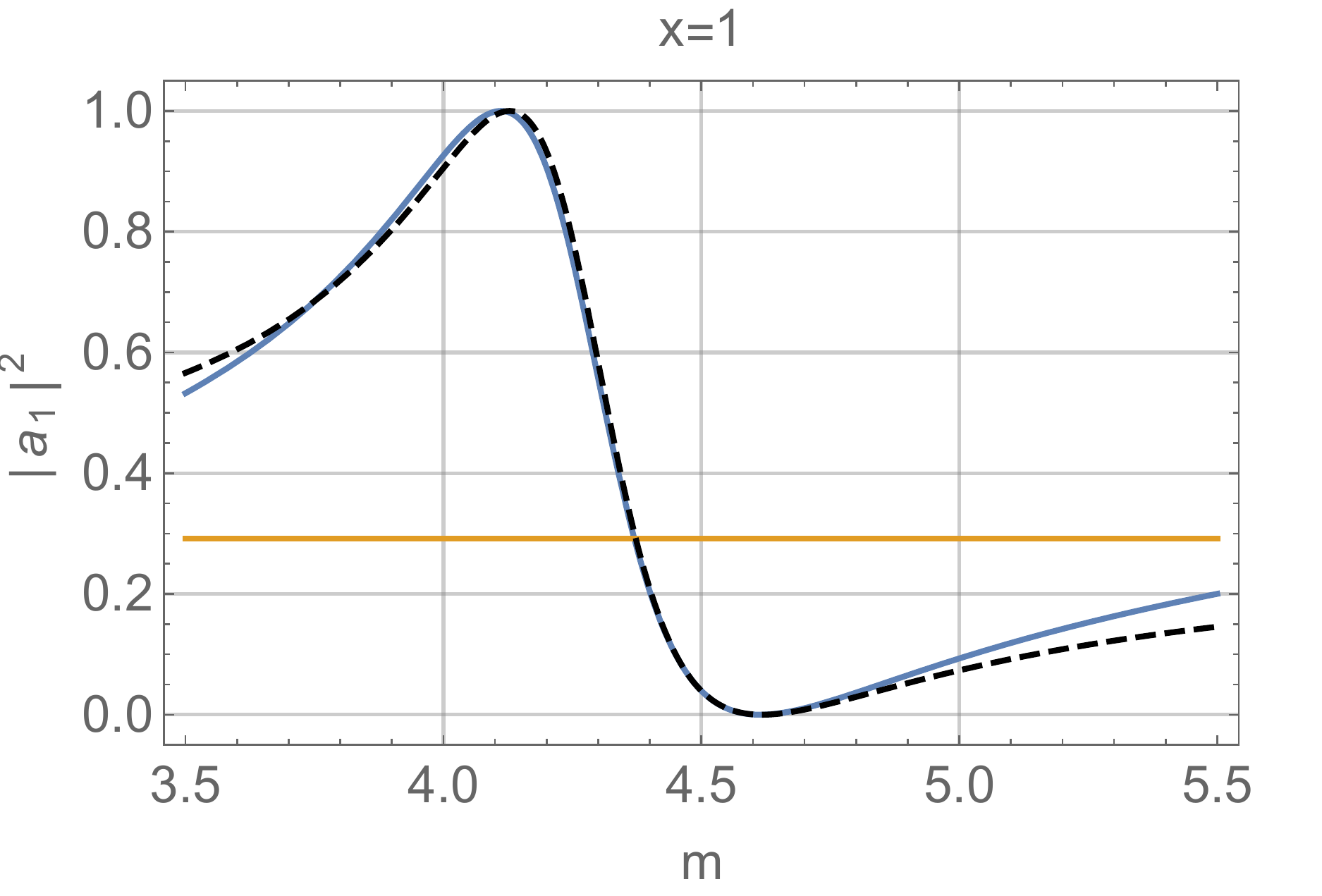}\\
  \caption{(color online) Comparison of profile $|a_1(m)|^2$, given by the exact Mie solution (blue solid line), with that, produced by Eqs.~\eqref{Eq:q_and_epsilon_a}, \eqref{Eq:|a_n|_Fano}; \eqref{Eq:epsilon_though_delta_m}(black dashed line); $x=1$. A yellow horizontal line corresponds to $a_1^{\rm (PRS)}$.}\label{fig:Fano_exact_vs_approx}
\end{figure}

\begin{figure}
  \centering
  \includegraphics[width=\linewidth]{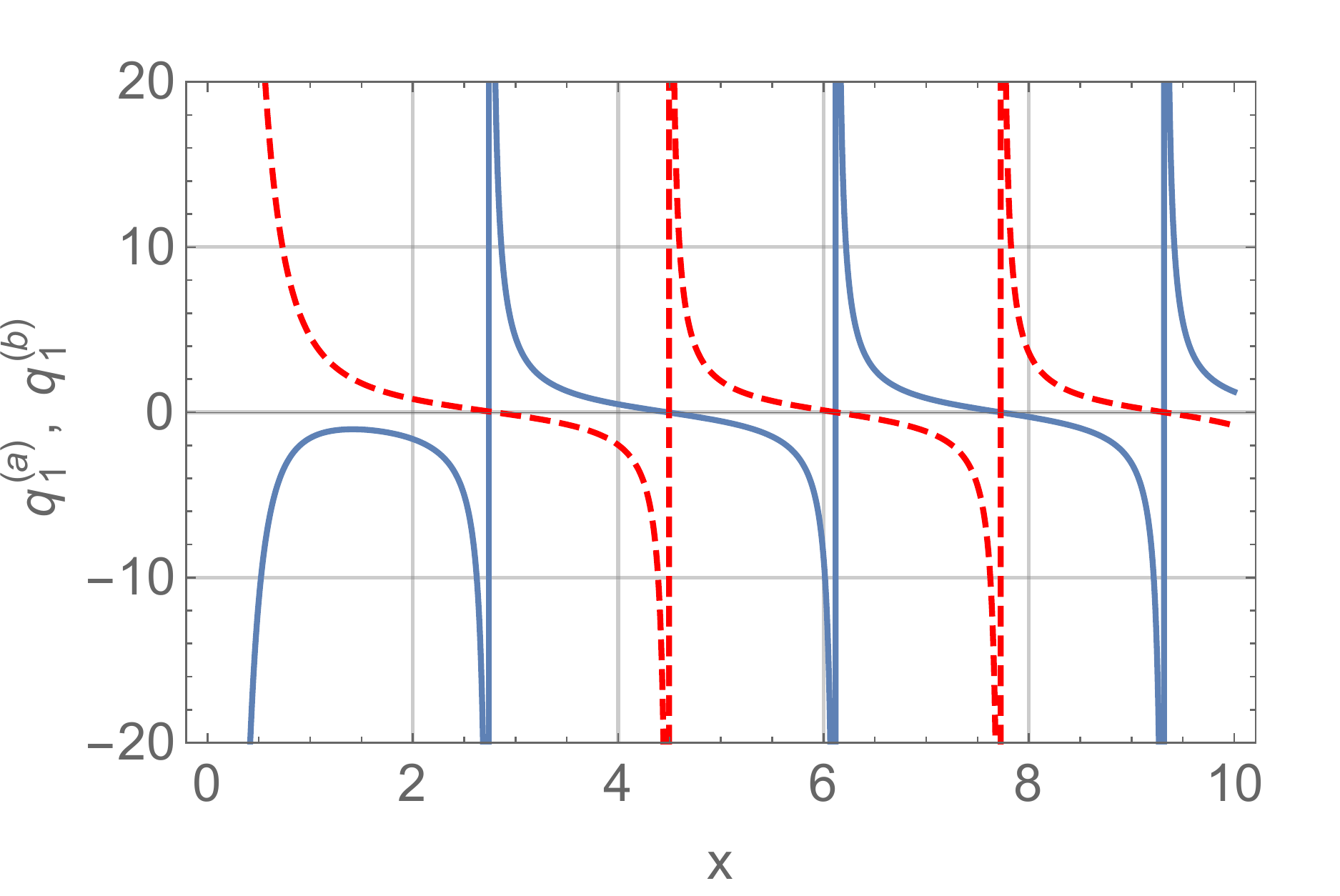}\\
  \caption{(color online) $q_1^{(a)}(x)$ -- solid and  $q_1^{(b)}(x)$ -- dashed according to Eqs.~\eqref{Eq:q_and_epsilon_a}, \eqref{Eq:qb_n}. }\label{fig:qab}
\end{figure}

Once again we encounter a discrepancy, this time with our own results. It seems that by means of just identical transformations we have proven that the profiles of $|a_n|^2,\;|b_n|^2$ are of the Fano type always. On the other hand, it has been shown in our previous publication~\cite{Tribelsky:PRL:2008} that at least in the case of small particles \mbox{($mx \ll 1$)} the Fano resonances in partial scattered coefficients cannot happen.

Naturally, the discrepancy, as usual, is an illusion. The fact is that the condition $mx \ll 1$ does not allow $m$ to be large enough to reach the point of the first Fano resonance. Therefore, to prove that the profile, Eq.~\eqref{Eq:|a_n|_Fano} does correspond to the Fano lineshape, we have to show that (i) in the specified range of variations of $m\;\;\epsilon_n^{(a)}(m)$ reaches the values corresponding to the constructive ($\epsilon_n^{(a)} = 1/q_n^{(a)}$) and destructive ($\epsilon_n^{(a)} = -q_n^{(a)}$) conditions, i.e., the problem in question may exhibit the Fano resonances, indeed, and (ii) for a given Fano profile in the vicinity of both the constructive and destructive interference $\epsilon$ is one and the same \emph{linear} function of $m$.
Regarding the former, the manifestation of the constructive and destructive resonances in the problem has been already shown in the previous sections of this paper and in other publications~\cite{Tribelsky:EPL:2012,Rybin:OE:2013,Rybin:SciRep:2014}. As for the latter, employing Eqs.~\eqref{Eq:Fn_res_a}, \eqref{Eq:Gn_res_a}, after simple calculations, it is possible to show that:
\begin{figure}
  \centering
   \begin{tabular}{c}
   \includegraphics[width=0.5\textwidth]{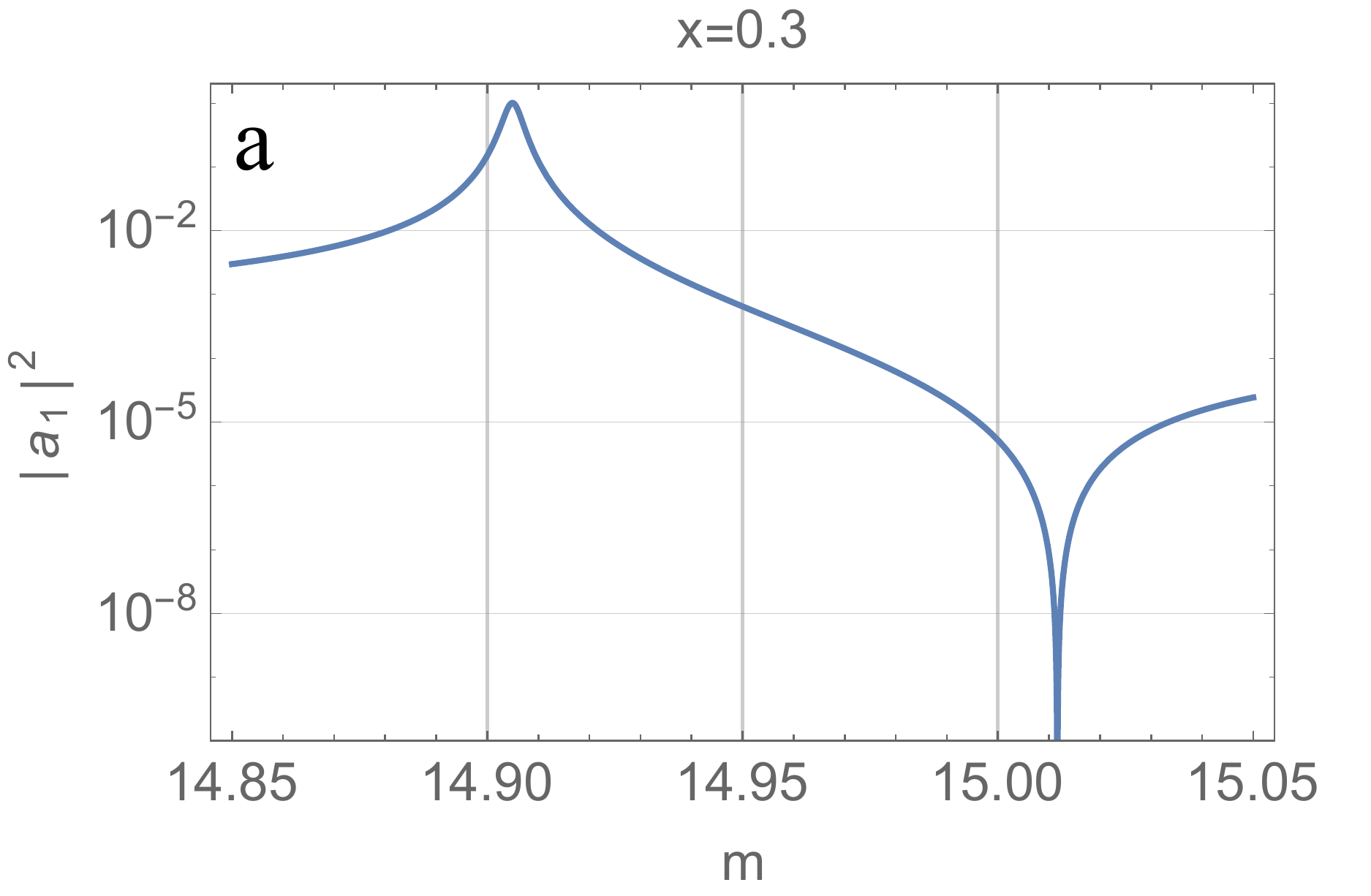}\\
   \includegraphics[width=0.5\textwidth]{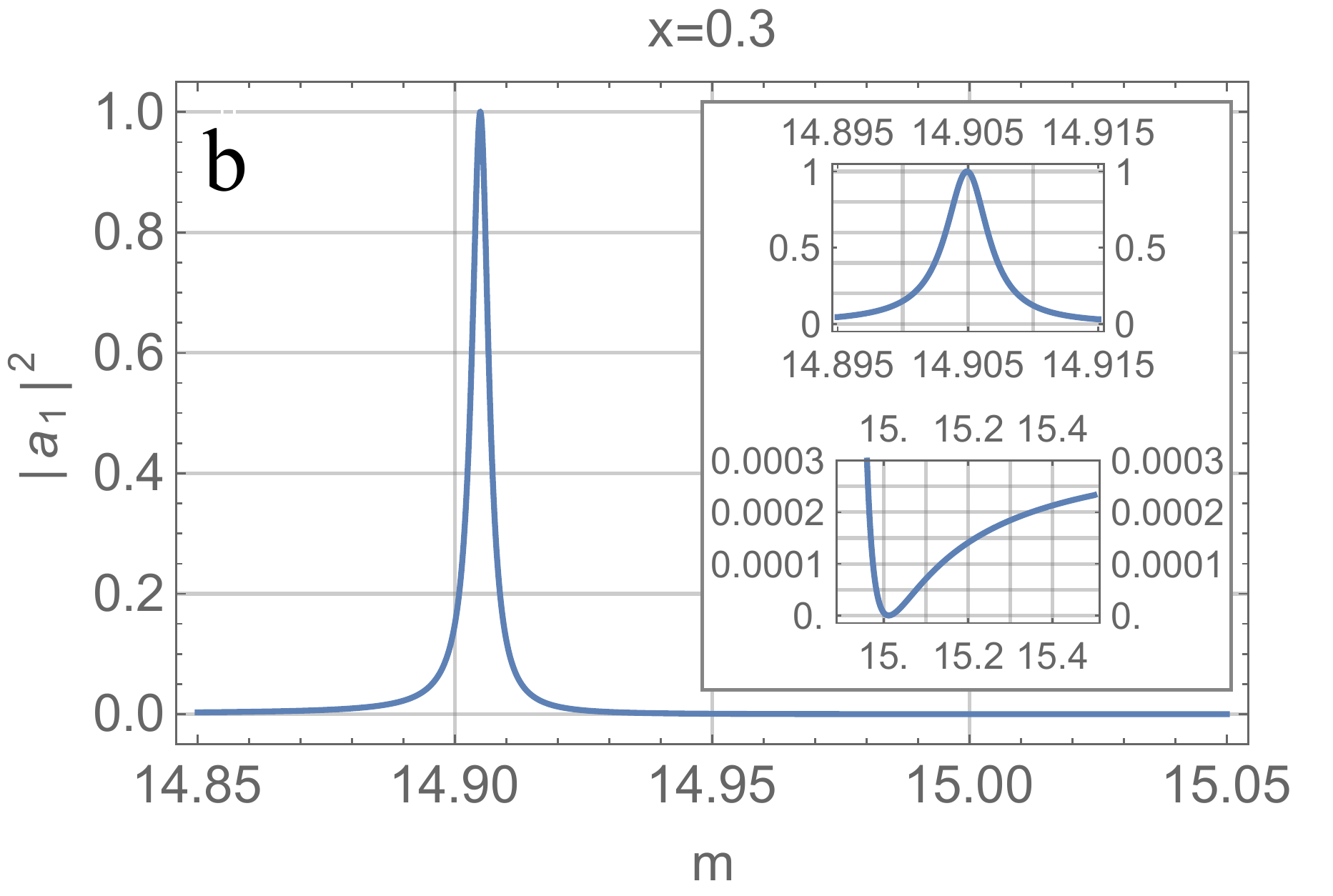}
   \end{tabular}
  \caption{(color online) An example of a reduction of the Fano profile to Lorentzian at $x=0.3$. While a logarithmic plot of $|a_1(m)|^2$ (a) clearly exhibits the Fano line with the point of the constructive resonance ($|a_1(m)|^2=1$) at $m \approx 14.905$ and the destructive resonance ($|a_1(m)|^2=0$) at $m \approx 15.012$, a linear plot of the same dependence (b) is, practically, indistinguishable from the Lorentzian profile, centered about the point of the constructive resonance. The high-resolution-lineshape in the vicinity of the constructive and destructive resonances is shown in the inset. The background level $|a_1^{{\rm (PRS)}}|^2 \approx 0.0003404$. It corresponds to $q^{(a)2} \approx 2937$.}\label{fig:Fano-Lorentz}
\end{figure}

\begin{eqnarray}
  \epsilon_n^{(a)} &=&  \left(\psi_n'^2(x) + \chi_n'^2(x)\right)m^{{\rm(res})}x\delta m\nonumber\\
  & & - \psi_n'(x)\psi_n(x) - \chi_n'(x)\chi_n(x).\label{Eq:epsilon_though_delta_m}
\end{eqnarray}
It means that in the vicinity of the resonances $\epsilon_n^{(a)}$ is a linear function of $\delta m$, indeed. Regarding the validity of this approximation up to the points of the extrema of the profile, it follows from Eqs.~\eqref{Eq:delta_m_G_a}, \eqref{Eq:delta_m_F_a}. Note that Eq~\eqref{Eq:epsilon_though_delta_m} yields that the characteristic scale for the Fano profiles in $\delta m$ is $1/m$, which agrees with the previous consideration of the linewidths, see Section \ref{sec:3}.

It is worthwhile mentioning that the accuracy of approximation, \eqref{Eq:epsilon_though_delta_m} is surprisingly good. As an example a comparison of presentation, Eqs.~\eqref{Eq:q_and_epsilon_a}, \eqref{Eq:|a_n|_Fano}, \eqref{Eq:epsilon_though_delta_m} with the exact Mie solution for $|a_1|^2$ at $x=1$ in the vicinity of $m=4.5$ is presented in Fig.~\ref{fig:Fano_exact_vs_approx}.

%

It is important that the asymmetry parameter $q_n^{(a,b)}$ in this case is expressed by simple formulas, Eqs.~\eqref{Eq:q_and_epsilon_a}, \eqref{Eq:qb_n} in terms of $\psi_n(x)$ and $\chi_n(x)$, i.e., it depends just on the multipolarity of the scattered partial wave $n$, radius of the scattering sphere and wavelength of the incident light (we remind that $x \equiv 2\pi R/\lambda$) and does not depend on the reflective index of the particle $m$.

Note also that the expressions for $q_n^{(a,b)}$ may be obtained in a less formal way too. According to the preceding analysis in the off-resonance regions \mbox{$a_n \cong a^{({\rm PRS})}$,} \mbox{$b_n \cong b^{({\rm PRS})}$.} In terms of the Fano resonances these regions correspond to the limit $\epsilon \rightarrow \infty$, when the Fano profile tends to $1/(1+q^2)$, see Eq.~\eqref{Eq:|a_n|_Fano}. Equalizing $a^{({\rm PRS})}$ to $1/(1+q^{(a)2})$ and $b^{({\rm PRS})}$ to $1/(1+q^{(b)2})$ we again arrive at Eqs.~\eqref{Eq:q_and_epsilon_a}, \eqref{Eq:qb_n} for $q_n^{(a,b)}$.

Let us discuss the dependence $q_n^{(a,b)}(x)$ in detail. At $x > n^2$ we just can take the first term of the asymptotical expansions of the Riccati-Bessel functions. For $\psi_n(x)$ it is given by Eq.~\eqref{Eq:sin=0}. The corresponding asymptotics for $\chi_n(x)$ reads~\cite{DLMF::2015}
\begin{equation}\label{Eq:chi_n_at_large_x}
  \chi_n(x) \cong \cos\left(x-\frac{n\pi}{2}\right).
\end{equation}
Then, at $x > n^2$
\begin{equation}\label{Eq:qab_at_large_x}
  q_n^{(a)} \cong - \tan\left(x-\frac{n\pi}{2}\right),\;\; q_n^{(b)} \cong \cot\left(x-\frac{n\pi}{2}\right).
\end{equation}
Note, Eq.~\eqref{Eq:qab_at_large_x} yields $q_{n}^{(a)}= q_{n \pm 1}^{(b)}=q_{n+2}^{(a)}$.

In the opposite limit of small $x$, utilizing the known asymptotic expressions for the Bessel functions at a small value of the argument, we arrive at the following expressions for $q_n^{(a,b)}$ at $x \ll 1$:
\begin{eqnarray}
 q_n^{(a)} & \cong & -\frac{n}{n+1} \frac{2^{1 + 2 n}\Gamma(n +\frac{1}{2}) \Gamma(n +\frac{3}{2})}{\pi x^{2n+1}}, \label{Eq:qa_at_small_x} \\
 & \mbox{}&\nonumber\\
 q_n^{(b)} & \cong & \frac{2^{1 + 2 n}\Gamma(n +\frac{1}{2}) \Gamma(n +\frac{3}{2})}{\pi x^{2n+1}}, \label{Eq:qb_at_small_x}
\end{eqnarray}
where $\Gamma(z)$ stands for the gamma function. As an example, the dependences $q_1^{(a,b)}(x)$ in domain $0\leq x \leq 10$, are shown in Fig.~\ref{fig:qab}.

Both expressions, Eqs.~\eqref{Eq:qa_at_small_x} and Eq.~\eqref{Eq:qb_at_small_x} diverge as $1/x^{2n+1}$ at $x \rightarrow \infty$. On the other hand, at large $q^2$ the Fano profile, Eq.~\eqref{Eq:|a_n|_Fano} is reduced to the Lorentzian one:
\begin{equation}\label{Fano_to_Lorentz}
  \frac{\left(\epsilon+q\right)^2}{\left(1+q^{2}\right)\left(1+\epsilon^{2}\right)} \xrightarrow[q \rightarrow \infty]{}\frac{1}{1+\epsilon^{2}}.
\end{equation}
\begin{figure*}
\centering
   \begin{tabular}{cc}
   \includegraphics[width=0.5\textwidth]{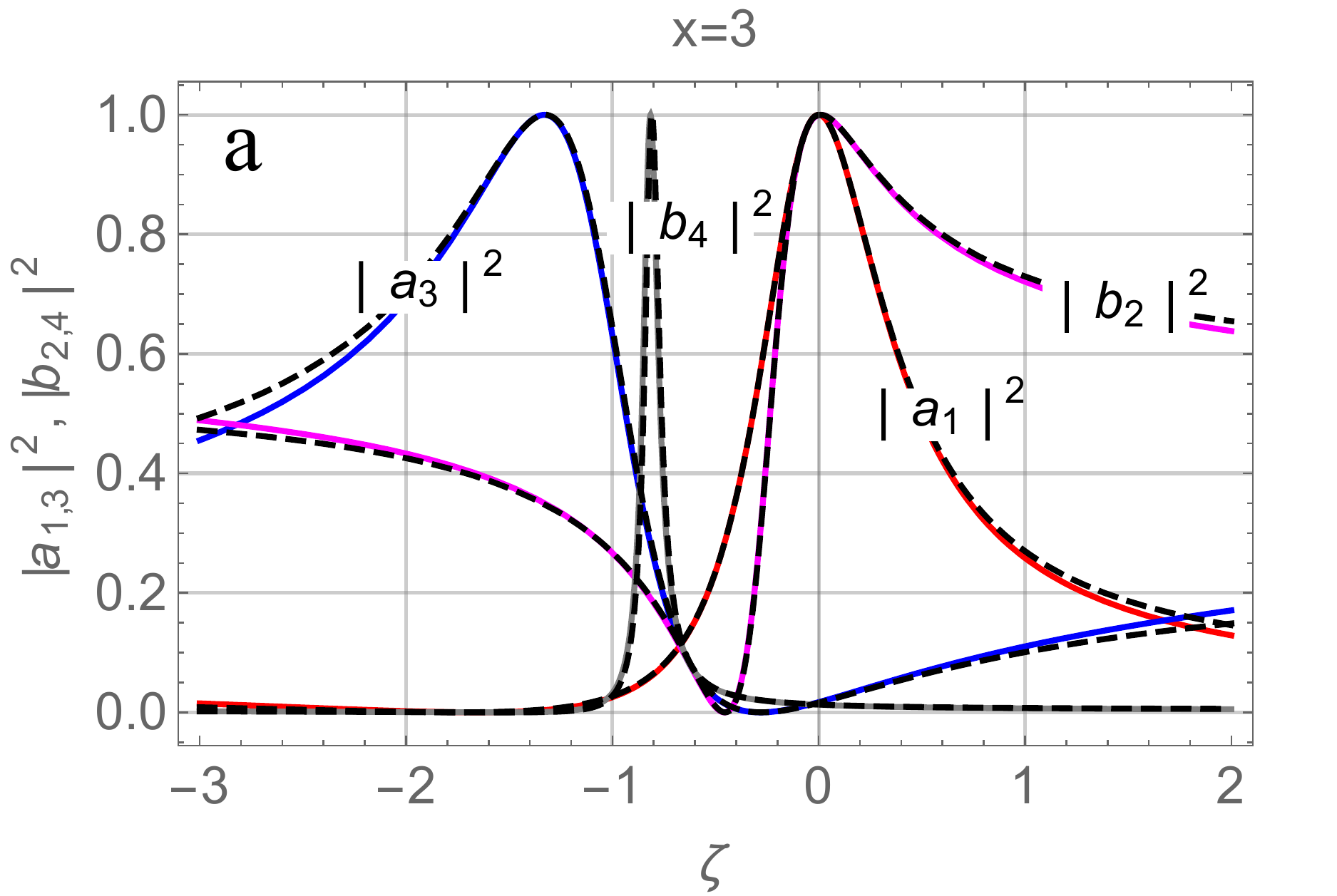}&
   \includegraphics[width=0.5\textwidth]{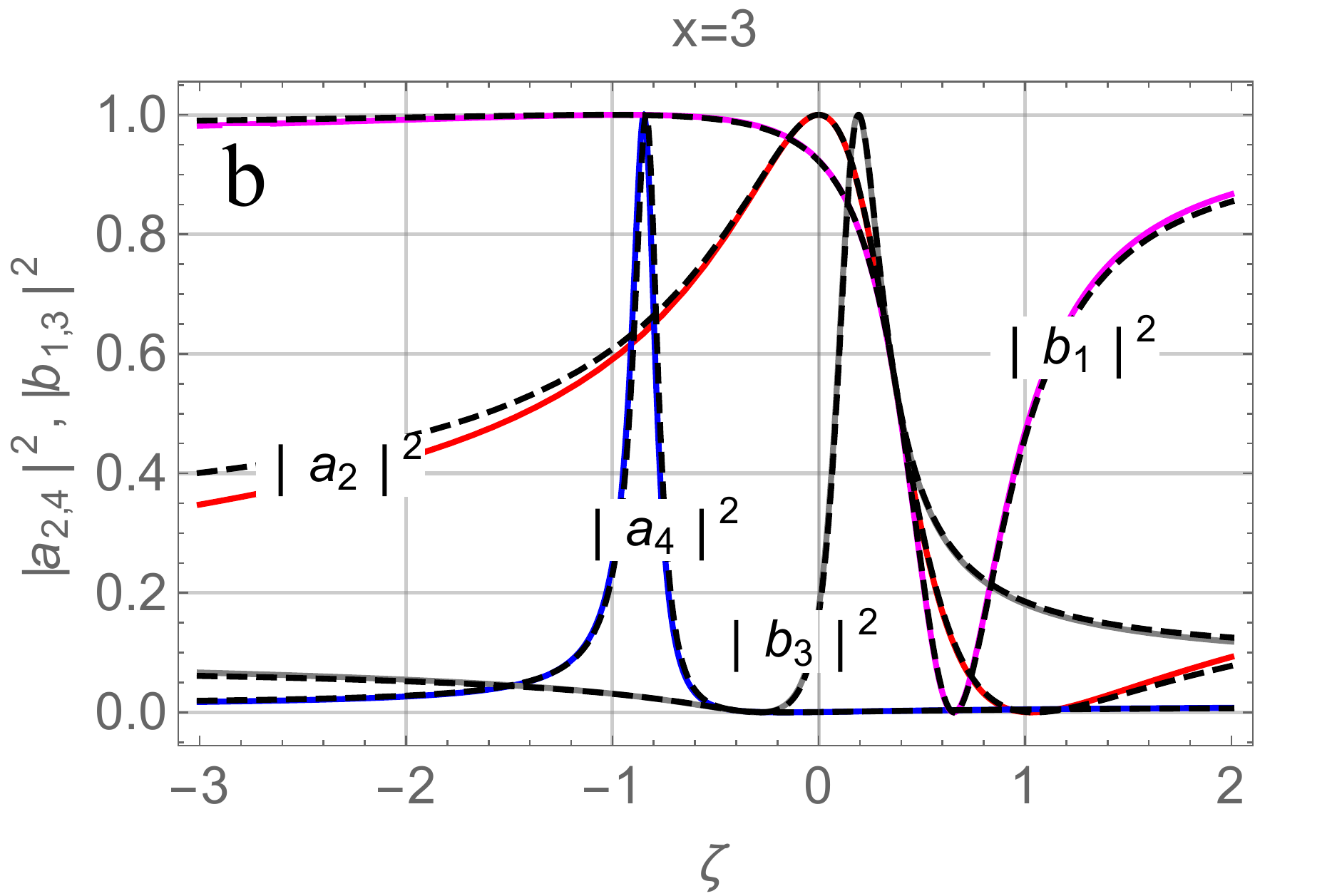}\\
   \includegraphics[width=0.5\textwidth]{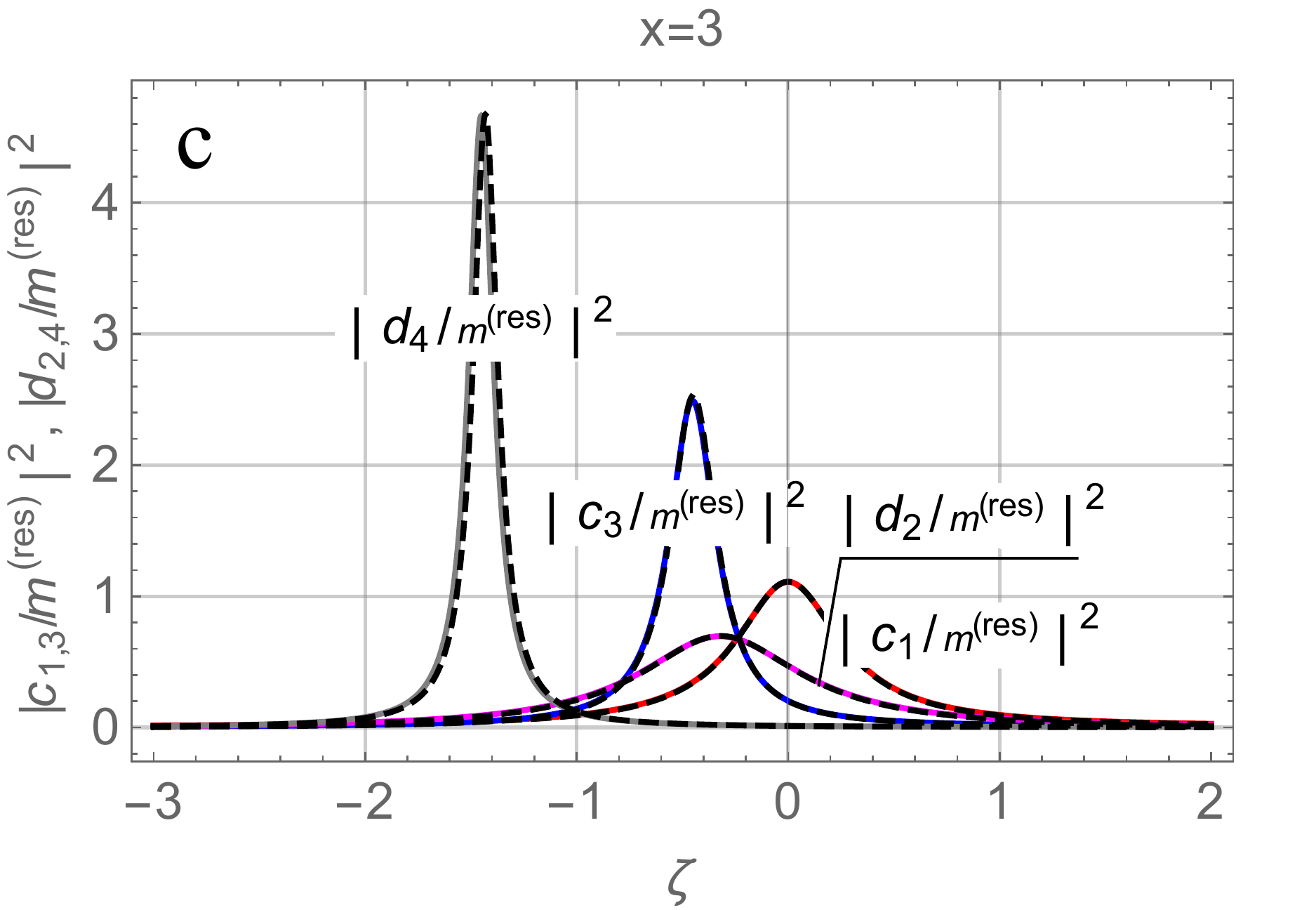}&
   \includegraphics[width=0.5\textwidth]{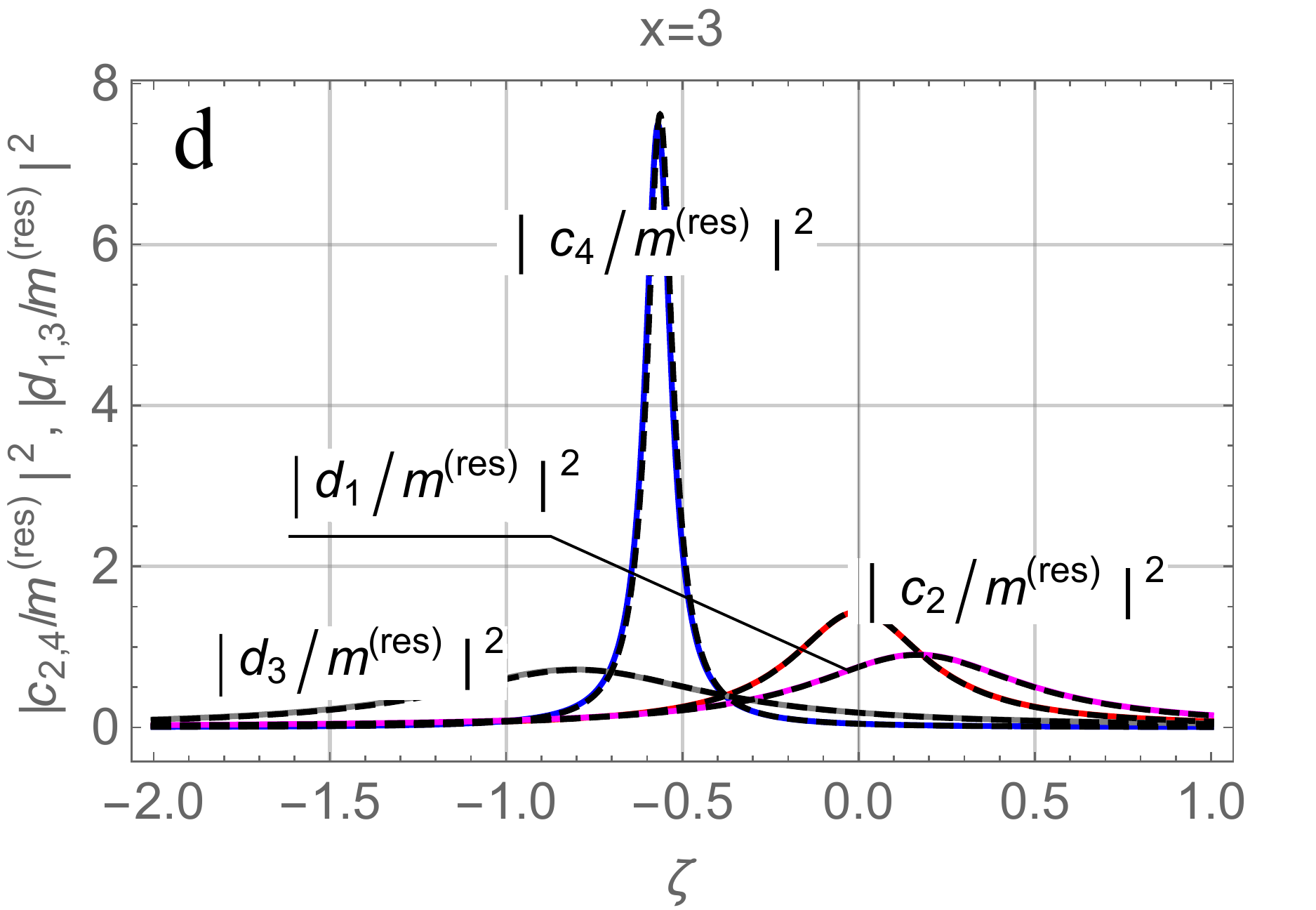}\\
   \end{tabular}
\caption{(color online) Collapse of lines of the resonances to a single universal set, as a result of the scale transformation, \mbox{Eqs.~\eqref{Eq:scale_m}, \eqref{Eq:scale_cd}} at $x =3;\;\zeta = (m - m^{({\rm res})})m^{({\rm res})}$: (a) -- color solid and black dotted lines correspond to $m^{({\rm res})} = 9.944...$ and $m^{({\rm res})} = 50.788...$, respectively (local maxima of $|a_1|^2$); (b) -- color solid and black dotted lines correspond to $m^{({\rm res})} = 9.350...$ and $m^{({\rm res})} = 50.252...$, respectively (local maxima of $|a_2|^2$); (c) -- color solid and black dotted lines correspond to $m^{({\rm res})} = 9.441...$ and $m^{({\rm res})} = 50.263...$, respectively (local maxima of $|c_1|^2$); (d) -- color solid and black dotted lines correspond to $m^{({\rm res})} = 9.919...$ and $\,m^{({\rm res})}\, = \,50.783...\,$, respectively (local maxima of $|c_2|^2$).}\label{fig:Fig4}
\end{figure*}

To understand the physical grounds for this reduction, we have to highlight the following. As it has been pointed out already, in the off-resonance regions the Fano profile tends to a non-zero constant background value, $1/(1+q^2)$, while the Lorentzian profile tends to zero, cf. the limits of the right- and left-hand-side of Eqs.~\eqref{Fano_to_Lorentz}. At the point of the constructive resonance ($\epsilon = 1/q$) the Fano profile exhibits its maximal value, equal to unity. So does the Lorentzian profile. However, if for the Lorentzian profile its increase from the off-resonance region also equals unity, for the Fano profile it equals the difference between unity and the background level, i.e., $1-[1/(1+q^2)] = q^2/(1+q^2)$. In the point of the destructive resonance the Fano profile drops from the background level to zero, that is to say, the amplitude of the corresponding modulation of the profile equals the background level $1/(1+q^2)$. The ratio of the amplitudes of the two modulations of the profile (constructive to destructive) is just $q^2$. The larger $q^2$, the less pronounced the destructive resonance relative to constructive and the closer the Fano profile to Lorentzian.

If now we remember, that the case $x \ll 1$ corresponds to a \emph{small} particle and a small particle in the off-resonance regions is a \emph{weak scatterer}, we immediately understand why the background level at $x \ll 1$ is small and the Fano profile is close to Lorentzian.

Thus, we have arrived at an important conclusion: \emph{though, formally, the Mie resonances for the partial scattered waves of the outer problem for small particles still belong to the Fano type, actually, the corresponding lineshape is very close to the Lorentzian}, cf.~\cite{Tribelsky:EPL:2012}. An example of such a profile is presented in Fig.~\ref{fig:Fano-Lorentz}.

\section{Scale invariance}

The results discussed in the previous sections give rise to a simple but very important conclusion. Namely, we have obtained that at any fixed $x$ and $n$ the profiles of $|a_n(m)|^2,\;|b_n(m)|^2$ exhibit infinite sequences of the Fano resonances. All these resonances have one and the same, $m$-independent value of $q$, see Eq.~\eqref{Eq:q_and_epsilon_a}, and the characteristic scale, decreasing as $1/m^{{(\rm res})}$, see Eqs.~\eqref{Eq:delta_m_G_a}, \eqref{Eq:delta_m_F_a}. Since the shape of the Fano profile is completely defined by the value of $q$, the latter means that at a given $x$ all the Fano resonances are identical and  \emph{ may be reduced to a single universal form} by the scale transformation
\begin{equation}\label{Eq:scale_m}
\delta m \rightarrow m^{({\rm res})}\delta m.
\end{equation}

Regarding $|c_n|^2,\;|d_n|^2$, at a fixed $m$ these profiles are Lorentzian, and, therefore, also universal. The corresponding scale transformations are Eq.~\eqref{Eq:scale_m}, supplemented with the rescaling of the coefficients
\begin{equation} \label{Eq:scale_cd}
  c_n \rightarrow c_n/m^{({\rm res})},\;\;  d_n \rightarrow d_n/m^{({\rm res})}.
\end{equation}

Finally, bearing in mind that the mismatches in the positions of the points of the resonances for modes with different $n$ are also scaled as $1/m^{({\rm res})}$, we obtain that the stipulated scale transformations should reduce the entire variety of the resonances to a single universal set, \emph{including the mutual position} of the resonances with different $n$. An example of such a collapse is shown in Fig.~\ref{fig:Fig4}.

It is also seen from Fig.~\ref{fig:Fig4} that an increase in $n$ results in shifts of the points of resonances for $|c_n|^2$ and $|d_n|^2$ to the left. For the Fano resonances (Fig.~\ref{fig:Fig4}a,b) the increase in $n$ is also accompanied by the change of the asymmetry parameter $q$, see Eq.~\eqref{Eq:q_and_epsilon_a}, and hence, by the change of the corresponding lineshape. Thus, to identify the shift one should look at the position of the local extrema. Note also, that for the lines $|c_n|^2,\;|d_n|^2$ the increase in $n$ results in an  increase in the $Q$-factor of the resonances, see Fig.~\ref{fig:Fig4}c,d. These peculiarities are generic for the problem in question and valid for any $n$.

\section{Dissipative effects}\label{sec:6}

Remember now, that the non-dissipative limit discussed above is a nonexistent in nature abstraction. In fact, dissipative losses always remain finite, as long as a real material is a concern. Then, a natural question ``How the obtained results are affected by the dissipative losses?" arises. In this section we are going to answer the question. To this end, we have to introduce a complex refractive index
\begin{equation}\label{Eq:mhat}
  \hat{m} = m + i\kappa
\end{equation}

Now note that, actually, there are two cases: strong dissipation ($m$ and $\kappa$ are of the same order of magnitude) and weak dissipation ($\kappa \ll m$). The former case is trivial --- the dissipation just suppresses the resonances. Thus, the most interesting is the weak-dissipation-case, especially its limit of the extremely weak dissipation ($\kappa \ll 1$), when the dissipative damping may compete with the small radiative one, and effects similar to the anomalous scattering~\cite{Tribelsky:PRL:2006} may be observed.

Examples of the extremely weak dissipation may be readily found among widely used semiconductors. For instance, at the wavelength of 532 nm (the second harmonic of a Nd:YAG laser) the complex refractive indices for gallium phosphide, silicon and gallium arsenide are \mbox{$\hat{m}_{_{\rm GaP}} = 3.4932 + i0.0026311;$} \mbox{$ \hat{m}_{_{\rm Si}} =  4.1360 + i0.010205;$} \mbox{$\hat{m}_{_{\rm GaAs}} = 4.1331 + i0.33609$,} respectively~\cite{refractiveindex.info}. In what follows all numerical examples, illustrating the developed theory, will be given for a particle made of gallium phosphate embedded into a transparent matrix with a refractive index close to unity. The dispersion of \mbox{$\hat{m}_{_{\rm GaP}}$} will be neglected, since in the proximity of the specified wavelength it is rather weak.

Let us discuss the weak dissipation case in detail. For this purpose, we have to consider small \emph{complex} departures of $\delta\hat{m} = \delta m + i\kappa$ from a purely real $m^{{(\rm res)}}$, defined by the condition \mbox{$\psi(m^{{(\rm res)}}x)=0$}. Expansion of $\psi(\hat{m}x)=0$ in powers of small $x\delta\hat{m}$ about the point $m^{{(\rm res)}}x$ results in the following trivial generalization of expression, Eq.~\eqref{Eq:d_n_tilde_AB} for  $|\tilde{d}_n|^2$:
\begin{widetext}
\begin{equation}\label{Eq:lineshape_|tilde_d_n|^2_with_kappa}
  |\tilde{d_n}|^2 =  \frac{1}{\left(m^{\rm (res)}x\delta m -A_n^{(d)}(x)\right)^2B_n^{(d)2}(x)
  +\left(1+\kappa B_n^{(d)}(x)m^{\rm (res)}x\right)^2},
\end{equation}
\end{widetext}
with the connection between $|\tilde{d}_n|^2$ and $|{d}_n|^2$
\begin{equation}\label{Eq:|dn|^2_through_|tilde_dn|^2}
  |d_n|^2 = \frac{m^2B_n^{(d)}(x)}{|\psi'_n(mx)|^2} |\tilde{d}_n|^2,
\end{equation}
following from Eqs.~\eqref{Eq:d_n_tilde}, \eqref{Eq:A_n^{(d)}_B_n^{(d)}}. Eq.~\eqref{Eq:lineshape_|tilde_d_n|^2_with_kappa} yields the linewidth
\begin{equation}\label{Eq:Gamma_|dn|^2_kappa}
  \gamma_{|d_n|^2}^{(\kappa)} = 2\frac{1+\kappa B_n^{(d)}(x)m^{\rm (res)}x}{B_n^{(d)}(x)m^{\rm (res)}x}.
\end{equation}

The profile $|d_n|^2$ is maximized by the same $\delta m_{|d_n|^2}^{\rm (res)}$, given by Eq.~\eqref{Eq:delta_m_|d|_max}, but the maximal value now is different:
\begin{equation}\label{Eq:Max_|dn|_kappa}
 \text{Max}\left\{ |d_n|^2\right\} \cong \frac{m^{{\rm (res)}2}B_n^{(d)}(x) }{|\psi'_n(m^{\rm(res)}x)|^2\left(1+\kappa B_n^{(d)}(x)m^{\rm (res)}x\right)^2}.
\end{equation}

The obtained results give rise to important conclusions. Namely, as it is clearly seen from Eq.~\eqref{Eq:Gamma_|dn|^2_kappa}, there is a point of crossover, $m_{\rm cr}$ from the non-dissipative regime (at $m \ll m_{\rm cr}$) to dissipative (at $m \gg m_{\rm cr}$), where $m_{\rm cr}$ is a solution of the equation:
\begin{equation}\label{Eq:m_cr}
   \kappa m x B_n^{(d)}(x) = 1.
\end{equation}
If at $m \ll m_{\rm cr}$ the linewidth is determined by Eq.~\eqref{Eq:linewidth_|d_n|^2}, at $m \gg m_{\rm cr}$ it converges to the universal $m$-$x$-$n$-independent value $2\kappa$.

However, the most dramatic changes happen with the amplitude of the resonances. While in the non-dissipative limit the amplitude of the profile $|d_n|^2$ at the resonance points at large $m$ increases as $m^2$, see Eq.~\eqref{Eq:d_n_through_d_n_tilde}, now the entire profile converges to the universal form
\begin{equation}\label{Eq:|dn|_kappa_large_m}
  |d_n|^2 \xrightarrow[{m \gg {m_{_{cr}}}}]{}\frac{1}{|\psi'_n(m^{\rm(res)}x)|^2x^2B_n^{(d)}(x)}\frac{1}{\delta m^2 + \kappa^2},
\end{equation}
which becomes completely $m^{\rm(res)}$-independent in the Fraunhofer regime, $m^{\rm(res)}x > n^2$ (we remind that in this regime $|\psi'_n(m^{\rm(res)}x)| \cong 1$).

Coefficients $c_n$ may be inspected exactly in the same manner. The corresponding expressions this inspection yields are quite similar to those, obtained above for $d_n$. For example,
\begin{eqnarray}
  & & \text{Max}\left\{ |c_n|^2\right\} \label{Eq:Max_|cn|_kappa}\\
  & &  \cong \frac{m^{{\rm (res)}2}B_n^{(c)}(x) }{|\psi_n(m^{\rm(res)}x)|^2\left(1-\kappa\frac{\psi''(m^{\rm (res)}x)}{\psi(m^{\rm (res)}x)} B_n^{(c)}(x)m^{\rm (res)}x\right)^2}.\nonumber
\end{eqnarray}
The reader should not be confused with sing minus in front of the term with the dissipative constant $\kappa$. The point it that this term includes also the factor ${\psi''(m^{\rm (res)}x)}/{\psi(m^{\rm (res)}x)}$, which is always negative and tends to minus unity in the Fraunhofer regime, see Eq.~\eqref{Eq:cos=0}. Thus, actually, the sign in front of the dissipative constant is plus. The other expressions for $c_n$, analogous to those discussed above for $d_n$, are not presented here owing to the triviality of the corresponding calculations. They result in the behavior of $c_n$ quite similar to that for $d_n$.

These peculiarities of $c_n$ and $d_n$ brings about a completely different scenario for vanishing of the Fano resonances for coefficients $a_n$ and $b_n$. For definiteness, let us focus on $a_n$. This coefficient is expressed in terms of $m$-independent $a^{{\rm (PRS)}}_n$ and $m$-dependent $d_n$ according to Eq.~\eqref{Eq:a_n_trough_aPRS_and_d_n}. Thus, the entire $m$-dependence of $a_n$ is given by the second term in the right-hand-side of Eq.~\eqref{Eq:a_n_trough_aPRS_and_d_n}. We remind the reader that in the non-dissipative limit $|a_n|^2$ always vanishes at the points of the destructive Fano resonances and reaches unity at the points of the constructive resonances, see Fig.~\ref{fig:a1} and Eq.~\eqref{Eq:|a_n|_Fano}. Then, the vanishing of the Fano resonances with an increase in $m$ occurs owing to the contraction of the resonance lines. Asymptotically the points of the constructive and destructive resonances merge, while \emph{the amplitude} of modulations of $|a_n|^2$ caused by the Fano resonances all the time \emph{remains the same}: each resonance forces $|a_n|^2$ to vary from zero to unity.
In contrast, now at large enough $m$ the width of the resonances for $|a_n|^2$ becomes $m$-independent and equal to $2\kappa$ (according to  Eq.~\eqref{Eq:a_n_trough_aPRS_and_d_n} the characteristic widths of the resonances for $a_n$ and $d_n$ have the same order of magnitude), while the amplitude of the modulations of $|a_n - a^{{\rm (PRS)}}_n|^2$ decreases as $1/m^2$, see Eqs.~\eqref{Eq:a_n_trough_aPRS_and_d_n},~\eqref{Eq:lineshape_|tilde_d_n|^2_with_kappa}. Thus, the vanishing of the Fano resonances (convergence of $a_n$ to $a^{{\rm (PRS)}}_n$) occurs owing to the vanishing of the \emph{amplitude} of the modulations of $a_n$. A crossover from the non-dissipative scenario to dissipative is again determined by $m_{{\rm cr}}$, see Eq.~\eqref{Eq:m_cr}. The behavior of $b_n$ is analogous.

\section{Resonances at varying size parameter and fixed refractive index}
\subsection{Amplitudes of resonances}

Up to now the resonances have been studied at a fixed $x$ (and $\kappa$) at varying $m$. On the other hand, as it has been mentioned above, the most interesting from the experimental viewpoint is the dependence of the resonances on the size of the particle at a fixed value of the refractive index. In this case at any fixed $n$ an increase in $x$ again results in a cascade of resonances, whose position is determined by the same conditions Eqs.~\eqref{Eq:psi=0}, \eqref{Eq:psi_prime=0}, regarded now as equations \mbox{for $x$.}

At $mx \gg 1$ these two problem formulations (fixed $x$ at varying $m$ and fixed $m$ at varying $x$) are easily reduced to each other by the set of transformations \mbox{$m^{{\rm (res)}} \leftrightarrow m$,} \mbox{$x \leftrightarrow x^{{\rm (res)}}$,} \mbox{$ x\delta m \leftrightarrow m\delta x$,} as long as the shape of a single resonance line is a concern, see the corresponding remark in the end of Sec.~\ref{sec:3}. However, this is not the case anymore, if we are interested in rather a large range of variations of the size parameter. The corresponding behavior of the scattering coefficients is discussed in the present section.
\begin{figure}
  \centering
  \begin{tabular}{c}
   \includegraphics[width=0.5\textwidth]{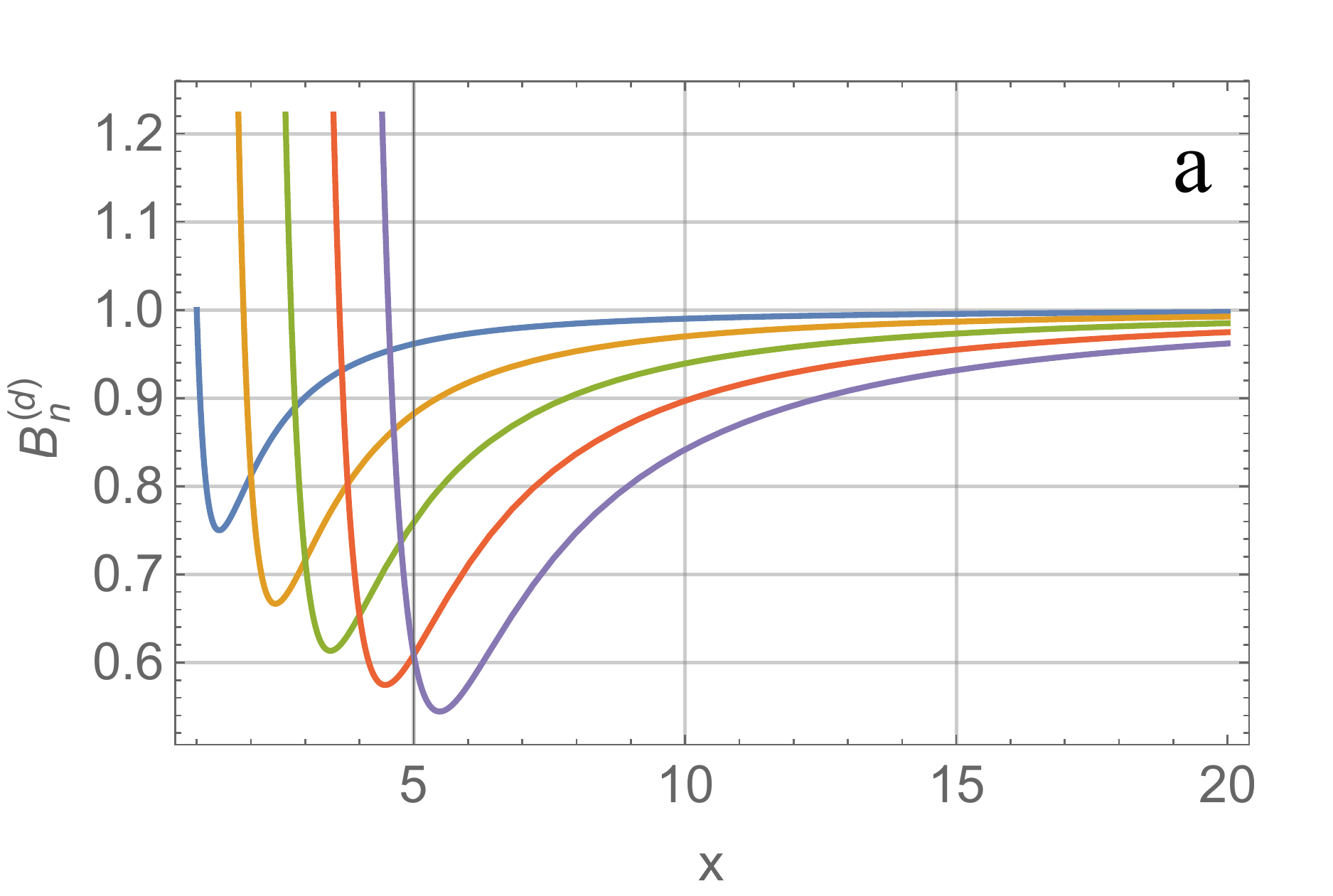}\\
   \includegraphics[width=0.5\textwidth]{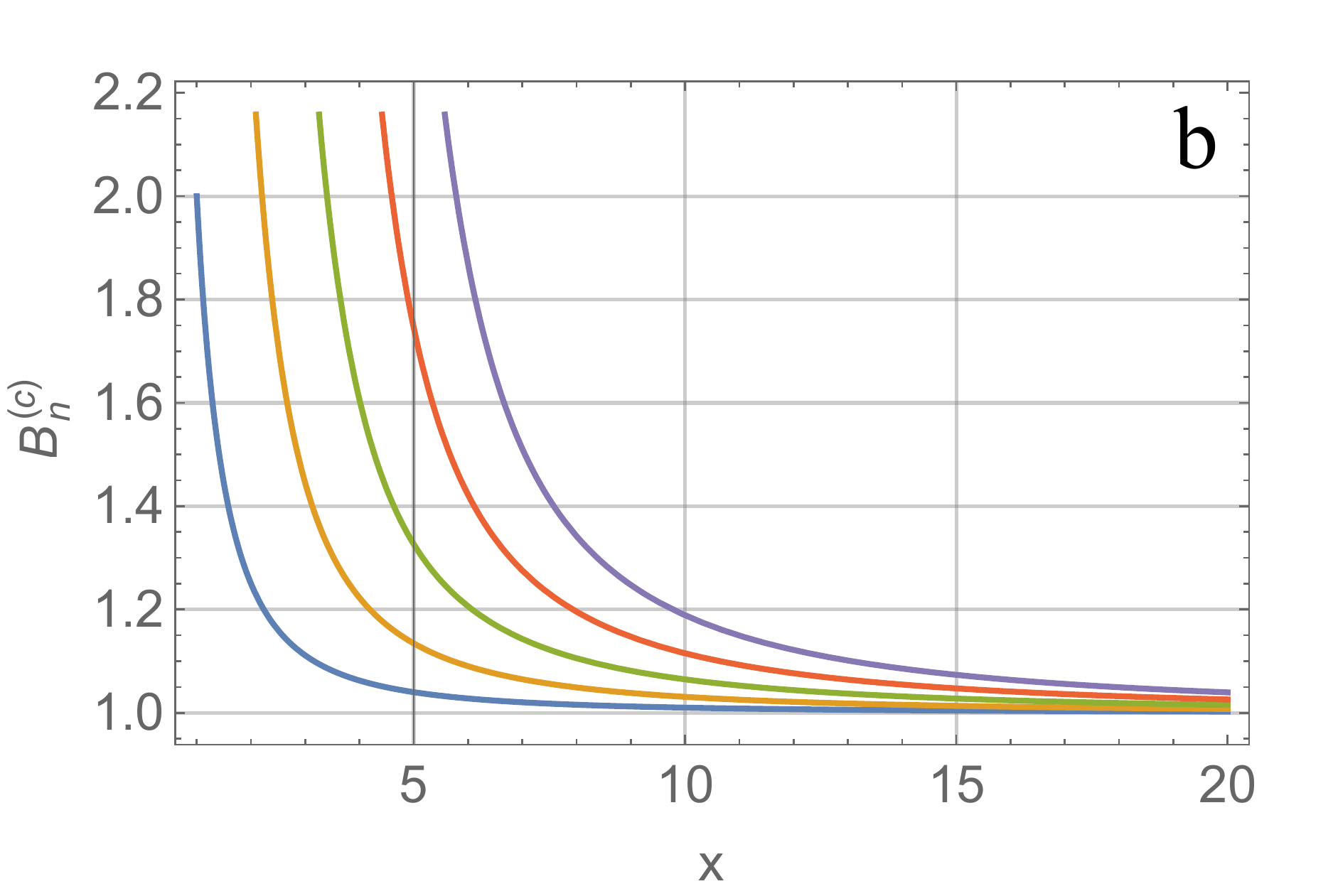}
   \end{tabular}
  \caption{(color online) The behavior of $B_n^{(d)}(x) \equiv |\xi_n'(x)|^2$ and $B_n^{(c)}(x) \equiv |\xi_n(x)|^2$ for the first five multipoles. The curves from left to right correspond to $n=1,\;2,\;3,\;4$ and 5, respectively.}\label{fig:Bn(x)}
\end{figure}

The results obtained in the previous Sec.~\ref{sec:6} show that the key function determining the behavior of the resonances at large variations of $x$, is $B_n^{(d)}(x) = |\xi_n'(x)|^2$. The known expansions of the Bessel functions in powers of their small argument yield the following behavior of $B_n^{(d)}(x)$ at $x \rightarrow 0$:
\begin{equation}\label{Eq:Bn_at_x->0}
  B_n^{(d)}(x) \cong \frac{n^2 4^n \Gamma^2(n+\frac{1}{2})}{\pi x^{2(n+1)}}.
\end{equation}
An increase in $x$ is supplemented with a monotonic decrease in $B_n^{(d)}$ until this function reaches its minimal value at $x \approx n$. A further increase in $x$ results in a slow monotonic growth of $B_n^{(d)}$, asymptotically approaching unity at $x \rightarrow \infty$, see Eqs.~\eqref{Eq:sin=0}, \eqref{Eq:chi_n_at_large_x}. Accordingly, the entire domain $0 \leq x < \infty$ of variations of $x$ is partitioned into two subdomains: the subdomain $0 < x \leq n$ of a sharp fall of $B_n^{(d)}(x)$ from infinity to its minimal value below unity and the one of a slow asymptotical growth $B_n^{(d)}(x)$ to unity $(n < x <\infty)$. As an example, functions $B_n^{(d)}(x)$ for the first five multipoles are presented in Fig.~\ref{fig:Bn(x)}a.
Thus, at any $n$ both $B_n^{(d)}(x)$ and $xB_n^{(d)}(x)$ are singular functions at $x\rightarrow 0$. It brings about a dramatic enhancement of the dissipative effects at small $x$, see Eq.~\eqref{Eq:Gamma_|dn|^2_kappa}. On the other hand, an increase in $B_n^{(d)}(x)$ increases the numerator in the right-hand-side of Eq.~\eqref{Eq:Max_|dn|_kappa}, defining the amplitude of the resonance. That is to say, at small $x$ there is a competition between the growth of the amplitude of the resonance owing to the increase in $B_n^{(d)}(x)$ in the numerator of Eq.~\eqref{Eq:Max_|dn|_kappa} and its suppression because of the growth of the same quantity in the denominator of the same equation. In this case we have to distinguish to limits:

(i) Despite the large value of $B_n^{(d)}$ the first (smallest) resonant $x^{({\rm res})}_{n,1}<n$ still corresponds to weak dissipation \mbox{($\kappa m x^{({\rm res})}_{n,1} B_n^{(d)}(x) \ll 1$),}. Then, the amplitude of this resonance is the largest for the given $n$. An increase in $x$ gives rise to the fall of the amplitudes of the sequential resonances, first due to the decrease of $B_n^{(d)}(x)$ in the numerator of Eq.~\ref{Eq:Max_|dn|_kappa}, and then due to the transition to the dissipation-controlled region due to increase of $\kappa B_n^{(d)}(x)m x$ in the denominator of this expression.

(ii) $x^{({\rm res})}_{n,1}<n$ is so small (i.e., $B_n^{(d)}(x^{\rm (res)}_{n,1})$ is so large), that $\kappa B_n^{(d)}(x^{\rm (res)}_{n,1})m x^{\rm (res)}_{n,1} \gg 1$, despite smallness of $\kappa$. Then,
the amplitude of the corresponding resonance approximately equals
\begin{equation*}
  \text{Max}\{|d_n|^2\} \cong \frac{1}{(\psi'_n(mx^{\rm (res)})\kappa x^{\rm (res)})^2B_n^{(d)}(x^{\rm (res)})},
\end{equation*}
see, Eq. \eqref{Eq:Max_|dn|_kappa}. Since, by definition, in a given cascade $x^{\rm (res)}_{n,1}<x^{\rm (res)}_{n,2}<x^{\rm (res)}_{n,3}<\ldots,$ the next resonances in the same cascade result in a decrease of $B_n^{(d)}(x^{\rm (res)}_{n,p})$ and, hence, in an \emph{increase} of their amplitude. It goes on in this manner until the decease of $B_n^{(d)}$, eventually, drives the particle out of the dissipation-controlled regime. Then, a further increase in $x^{\rm (res)}$ results in the effects, described above in \mbox{item (i).} Thus, now we have two dissipation-controlled domains: the first at small $x$ and the second at large, separated by a non-dissipative domain. The maximal amplitude of the resonance mode is achieved at $x^{\rm (res)}$ situated at the boundary between the first dissipation-control domain the non-dissipative domain.

\begin{figure}
  \centering
  \centering
  \begin{tabular}{c}
   \includegraphics[width=0.5\textwidth]{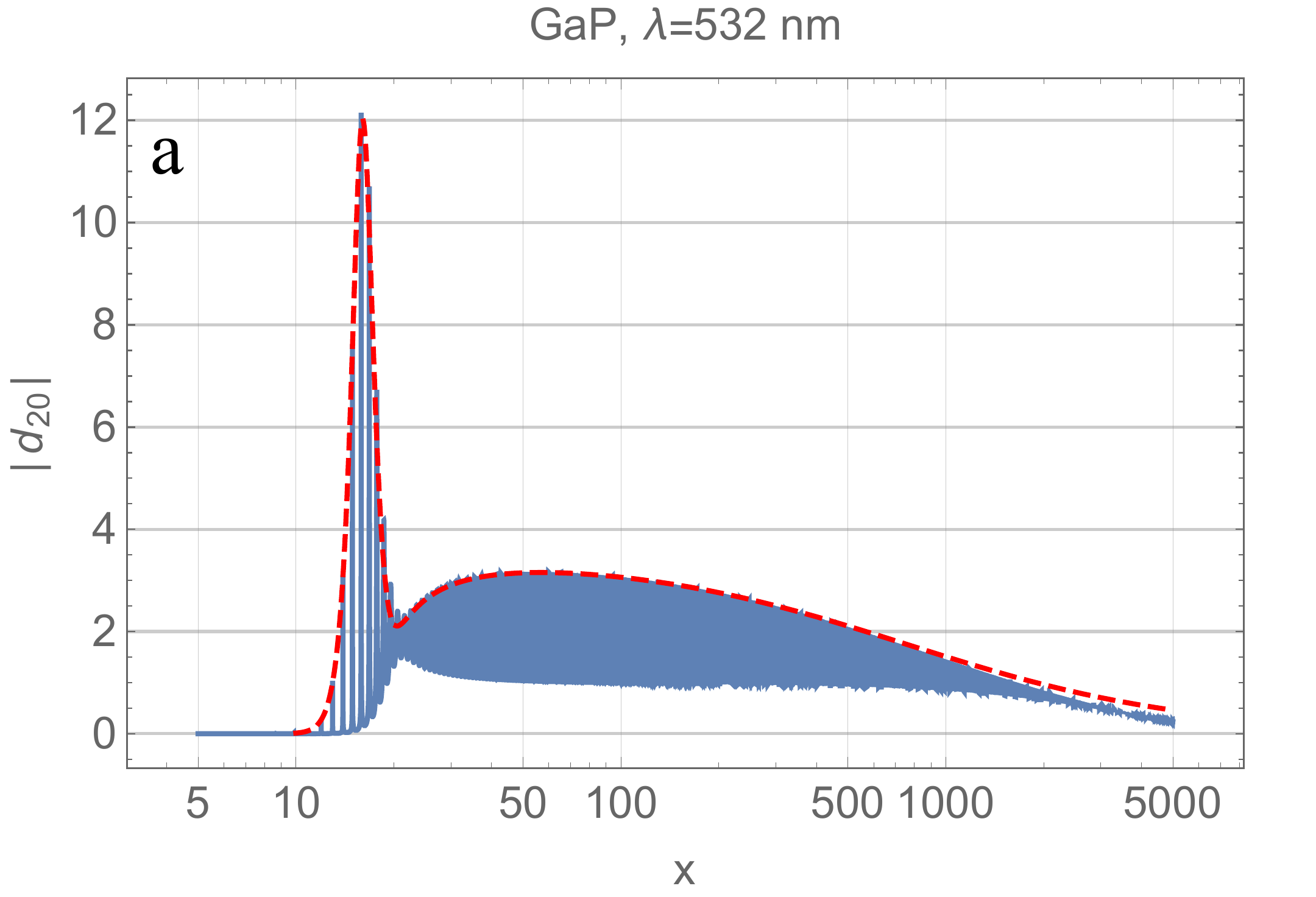}\\
   \includegraphics[width=0.5\textwidth]{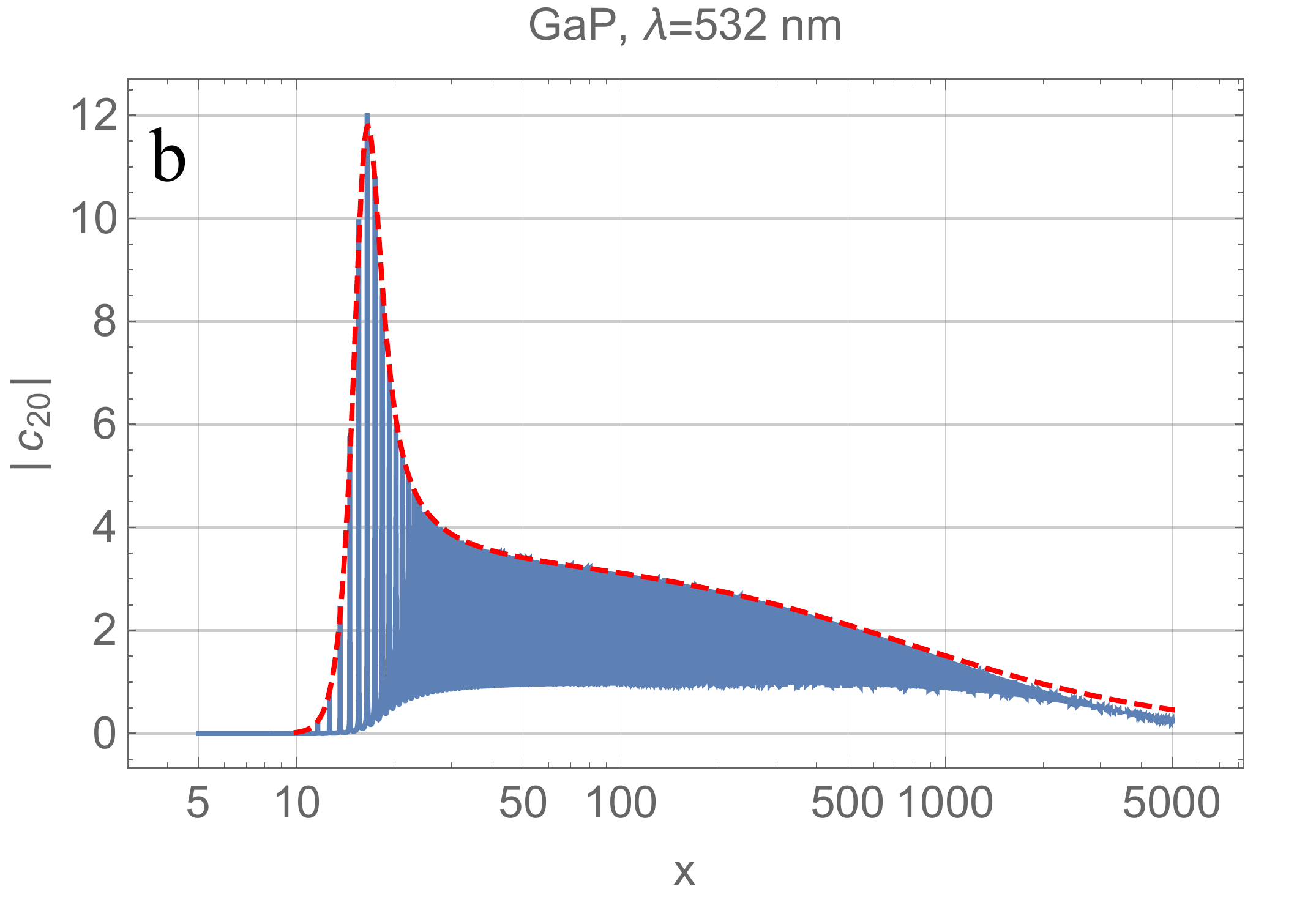}
   \end{tabular}
  \caption{(color online) A typical behavior of $|d_n(x)|$ and $|c_n|$ on the example of $n=20$ for a particle made of GaP. Blue solid lines corresponds to the exact solution, Eqs.~\eqref{Eq:d_n}~\eqref{Eq:c_n}; the envelopes, calculated according to the approximate Eq.~\eqref{Eq:Max_|dn|_kappa}~\eqref{Eq:Max_|cn|_kappa} are shown as red dashed lines. Note strongly non-monotonic dependence of the amplitude of the resonant oscillations on $x$ due to the enhancement of the dissipative effects at certain domains of variations of $x$ and a ``bottle neck" in the dependence $|d_n(x)|$ at the proximity of $x=n=20$. The bottle neck is originated in the corresponding minimum in $B_n^{(d)}(x)$, which does not have $B_n^{(c)}(x)$. For more details, see the text.   }\label{fig:GaP_d20}
\end{figure}

It is important to stress also the quite different asymptotic  behavior of profiles $|d_n(m)|^2$ at a fixed $x$, and $|d_n(x)|^2$ at a fixed $m$, respectively. If $|d_n(m)|^2$ at $m \rightarrow \infty$ converges to a certain universal periodic function, see Eq.~\eqref{Eq:|dn|_kappa_large_m}, $|d_n(m)|^2$ at $x \rightarrow \infty$ vanishes as $1/(\kappa x)^2$, owing to  Eq.~\eqref{Eq:Max_|dn|_kappa} and limits
\begin{equation*}
  |\psi'_n(mx^{\rm (res)})| \xrightarrow[{x^{\rm (res)} \to \infty }]{}1,\;\; B_n^{(d)}(x) \xrightarrow[{x \to \infty }]{}1.
\end{equation*}
The behavior of $|c_n|^2$ in general is analogous to the discussed above for $|d_n|$. However, in contrast to $B_n^{(d)}(x)$, function $B_n^{(c)}(x)$ is monotonically decreasing, see Fig.~\ref{fig:Bn(x)}b. This difference in $B_n$ results in a certain difference in the shape of the envelopes of the resonances for $|d_n(x)|$ and $|c_n(x)|$, see Fig.~\ref{fig:GaP_d20}.

\subsection{Overlap of resonances}

One of the key features of the problem in question is an overlap of a large number of resonances, which may result in a creation within the particle ``hot spots" with a giant concentration of the electromagnetic field. The corresponding field structure is determined by the coordinate dependence of the resonant modes. Its detailed inspection requires a separate consideration and will be reported elsewhere. In the present paper we discuss just the necessary conditions for such hot spots to come into being, namely the overlap of profiles of the modula of the scattering coefficients, describing the field within the particle. As usual, for the sake of briefness we restrict the discussion with the properties of $|d_n|^2$ solely. The behavior of $|c_n|^2$ may be studied in the same manner and exhibits similar peculiarities.

Overlap of two resonances means that the mismatch between the positions of the maxima of their lines is smaller than the largest linewidth. Therefore, to begin with, we have to discuss the linewidths of the resonances at a fixed $m$ and varying $x$. To this end, according to the stipulated above general rules, we have to replace in Eq.~\eqref{Eq:lineshape_|tilde_d_n|^2_with_kappa} $x\delta m \rightarrow m\delta x,\; m^{\rm (res)} \rightarrow m$ and $x \rightarrow x^{\rm (res)}$. This brings about the following expression for the linewidth:
\begin{equation}\label{Eq:gamma_x}
   \gamma_{|d_n|^2}^{(\kappa,x)} = 2\frac{1+\kappa B_n^{(d)}(x^{\rm (res)})mx^{\rm (res)}}{B_n^{(d)}(x^{\rm (res)})m^2}.
\end{equation}
Thus, in the non-dissipative region ($\kappa B_n^{(d)}(x)mx \ll 1$) the linewidth is $2/[B_n^{(d)}(x)m^2]$, while in the dissipation-controlled region ($\kappa B_n^{(d)}(x)mx \gg 1$) it is $2\kappa x/m$.

On the other hand, the resonant values of $x$ are defined by the condition $\psi_n(mx^{{\rm (res,}E)}) = 0$ for the electric modes and \mbox{$\psi'_n(mx^{{\rm (res,}H)}) = 0$} for magnetic, see Eqs.~\eqref{Eq:psi=0}, \eqref{Eq:psi_prime=0}. Below the Fraunhofer regime, at $mx < n^2$ different solutions of these equations are situated at distances of the order of $1/m$. It is always much larger than the linewidth in the non-dissipative region and still larger than that in the dissipation-controlled case, provided $x$ is of the order of unity, or smaller. Therefore, the overlap of the resonances in these regions may occur just accidentally.

However, the case is changed drastically in the Fraunhofer regime, when $mx$ becomes larger than $n^2$. In this regime points of different resonances are going to merge. In the leading (in $1/mx$) approximation they are just coincide, see Sec. \ref{sec:2}. To resolve the mismatch between the points of different resonances we have to go beyond the leading approximation. Employing the asymptotic expansion of the Riccati-Bessel functions at large values of their argument~\cite{DLMF::2015} and taking into account the first subleading term, instead of Eq.~\eqref{Eq:sin=0} we arrive at the following equation, determining the points of the electric mode resonances:
\begin{equation}\label{Eq:sin=0_subleading}
  \sin\left(\rho_{_E}-\frac{n\pi}{2}\right) + \frac{n(n+1)}{2\rho_{_E}}\cos\left(\rho_{_E}-\frac{n\pi}{2}\right) \cong 0,
\end{equation}
where $\rho_{_E} > n^2$ stands for $mx^{{\rm (res,}E)}$. Looking for a solution of this equation in the form $\rho_{_E} = \rho_{_E}^{(0)} + \delta \rho_{_E};$ $|\delta\rho_{_E}| \ll 1$, where $\rho_{_E}^{(0)}=[(2p+n)\pi/2]$ it is easy to obtain that
\begin{equation}\label{Eq:delta_rho_E}
  \delta\rho_{_E} \cong - \frac{n(n+1)}{\pi(n+2p)}.
\end{equation}
The solution is valid at $n(n+1) \ll \pi(n+2p)$. The corresponding mismatch between different resonances is $\delta \rho_{_E}/m$. Strictly speaking, we have to add to this mismatch another one, caused by the departure of the position of the maxima of $|d_n|^2$ from the resonance points determined according to the employed condition $\psi_n(mx) =0$. However, according to Eq.~\eqref{Eq:lineshape_|tilde_d_n|^2_with_kappa} this mismatch equals $A_n^{(d)}(x)/m^2$ (we recall, that to apply Eq.~\eqref{Eq:lineshape_|tilde_d_n|^2_with_kappa} to our case we have to replace $x\delta m \to m\delta x$). At large $m$ this quantity is small relative to $\delta\rho_{_E}/m$ and may be neglected.

\begin{figure}
  \centering
   \includegraphics[width=0.5\textwidth]{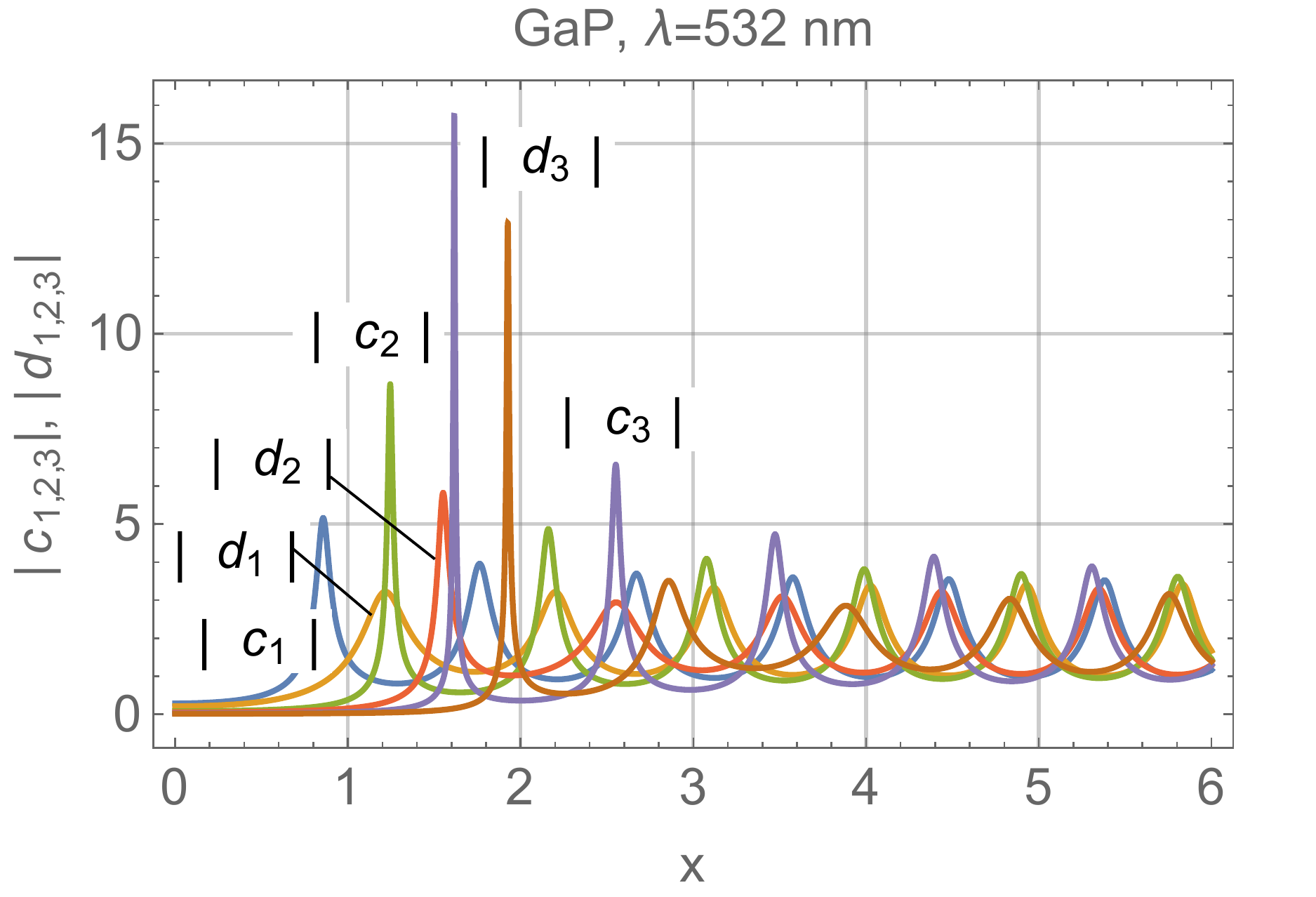}
  \caption{(color online) Cascades of resonances for $|c_n|$ and $|d_n|$ at a fixed refractive index and varying size parameter, $x$ for a spherical particle made of GaP, embedded in a transparent medium with the refractive index close to unity. Calculations according to Eqs.~\eqref{Eq:d_n},~\eqref{Eq:c_n}. Gradual overlap of resonances for \mbox{$|c_1|$ -- $|d_2|$ -- $|c_3|$} and \mbox{$|d_1|$ -- $|c_2|$ -- $|d_3|$} at an increase in $x$ is seen clearly. Note broadening of the resonance lines with an increase in $x$ and an increase of the amplitudes of the first resonances in each cascade with an increase in $n$. The first resonances in each cascades have the largest amplitudes. The large mismatches between the first resonances in the cascades for $|c_1|$ -- $|d_2|$ and $|c_2|$ -- $|d_3|$ are generic, while the overlaps of the first resonances in the cascades for $|d_1|$ -- $|c_2|$ and $|d_2|$ -- $|c_3|$ are accidental. For more details, see the text.}\label{fig:GaP_ovlp}
\end{figure}
Similar inspection of $|c_n|^2$ gives rise to $\rho_{_H} = \rho_{_H}^{(0)} + \delta \rho_{_H}$ with
\begin{equation}\label{Eq:delta_rho_H}
 \rho_{_H}^{(0)} = \frac{(n+2p+1)\pi}{2},\;\; \delta\rho_{_H} \cong \frac{n(n+1)}{\pi(n+2p+1)}.
\end{equation}

Let us try to collect all together. According to the results discussed in Sec.~\ref{sec:2}, there are two types of possible overlaps. First, corresponds to the overlap of the modes of the same type (e.g., electric -- electric, or magnetic -- magnetic) with different $n$ and $p$, changed in such a manner that sum $n+2p$ remains fixed. The adjacent candidates for this overlap have the difference in $n$ equals 2 and the difference in $p$ equals 1. According to Eqs.~\eqref{Eq:delta_rho_E}, \eqref{Eq:delta_rho_H} the mismatches between the resonance points for these two modes are
\begin{equation}\label{Delta_x_n_n+2}
  (\Delta x_{_E})_{n,n+2} \cong (\Delta x_{_H})_{n,n+2} \cong \frac{2(3+2n)}{m\pi N},
\end{equation}
where integer $N$ satisfies the condition $\pi N/2 = \rho_{_H}^{(0)}=\rho_{_H}^{(0)} = mx^{\rm (res)}$.

The candidates for the second type of the overlap are modes with the different nature: electric -- magnetic. In this case the adjacent modes have both $n$ and $p$ differed in unity, however, the signs of these mismatches are opposite, see Eqs.~\eqref{Eq:delta_rho_E}, \eqref{Eq:delta_rho_H}. Then, the corresponding mismatch is
\begin{equation}\label{Delta_x_EH_n_n+1}
    (\Delta x_{_{E,H}})_{n,n+1} \cong \frac{2(n+1)^2}{m\pi N}.
\end{equation}

In all the cases the mismatches monotonically increase with an increase in $n$. A rough estimate of the number of modes, which may overlap yields from the applicability condition of the Fraunhofer regime: $\rho > n^2$. For a given $x^{\rm (res)}$ it results in the inequality $n < \sqrt{N} \sim \sqrt{mx^{\rm (res)}}$.

Thus, at $n \gg 1$ the mismatches between the modes of the same type are scaled as \mbox{$n/(mN) \sim n/(m^2x)$,} while the ones between the modes of different types are scaled as $n^2/(mN)$. On the other hand, the linewidth in the dissipation-controlled region is scaled as $\kappa x/m$, see Eq.~\eqref{Eq:gamma_x}. An increase in $x$ decreases $\Delta x$ and increases the linewidth. The overlap begins, when the later becomes larger than the former. First, it happens with the modes of the same type at $x = x_{\rm ovlp}$, where
\begin{equation}\label{Eq:x_ovlp}
  x_{\rm ovlp} \sim \sqrt{\frac{n}{m\kappa}}.
\end{equation}

A further growth of $x$ increases the number of the overlapping modes, until all modes lying in the Fraunhofer range overlap. It happens at $x>X_{\rm ovlp}$, where $X_{\rm ovlp}$ is such a value of $x$, when $(\Delta x)_{1,\sqrt{N}}$ becomes equal, or smaller than the linewidth. Here $(\Delta x)_{1,\sqrt{N}}$ stands for the mismatch between the mode with $n=1$ and the mode with the maximal $n$ still belonging to the Fraunhofer range, i.e., with $n \sim \sqrt{N}$. According to Eqs. \eqref{Eq:delta_rho_E}, \eqref{Eq:delta_rho_H}, $(\Delta x)_{1,\sqrt{N}} \sim (n^2/mN)_{_{n=\sqrt{N}}} = 1/m$. It yields a simple estimate $X_{\rm ovlp} \sim 1/\kappa$.

As it follows from Eq.~\eqref{Eq:Max_|dn|_kappa} Max$\{|d_n(X_{\rm ovlp})|\} \sim 1$. Then, in the point, where all the overlapping modes are summarized coherently with the weight factor of the order of unity, the resulting amplitude of the electric (magnetic) field is $\sqrt{N}$ times greater (the density of energy N times greater) the one in the incident wave.

To give a certain impression about a manifestation of the discussed resonances in a possible real experiment we produce some results for a particle made of gallium phosphide. The corresponding plots are presented in Figs.~\ref{fig:GaP_d20},~\ref{fig:GaP_ovlp}.

Our calculations for GaP show also that the absolute maximum for a single resonant mode is achieved for the first resonance of mode $c_5$ at $x \approx 2.318$. The corresponding value of $|c_5|$ equals 34.641. The first resonant of $|d_5|$ gives rise to the close value: 33.161 at $x \approx 2.626$. The density of the electromagnetic energy in the hot spots corresponding to these solitary resonances is about $10^3$ greater than that for the incident wave. Meanwhile, multiple overlap of not so sharp resonances for a particle with $x \approx 25$ may produce hot spots with the density of energy $10^4$ relative to that in the incident wave.



\section{Conclusions}
Summarizing the results obtained, we may say that the detailed study of light scattering by a high refractive index particle with low dissipation discussed in the present paper, has revealed a number of new important features of the problem. In particular, we have shown that while at increasing $m$ the partial scattered waves outside the particle at any fixed $x$ and $n$ tend to the limits, corresponding to the light scattering by a perfectly reflective sphere, the field within the particle does not have any limit at all.

The reason for this difference in the behavior between the solutions of the outer and inner problems is related to the different manifestation of the infinite sequences of cascades of the Mie resonances in the two problems. For the scattered field outside the particle each Mie resonance in a cascade has the asymmetric Fano lineshape. As $m$ increases the resonances are suppressed, and the expressions for the scattering coefficients converge to the corresponding $m$-independent quantities for the perfectly reflecting sphere.

For the field within the particle, each Mie resonance in a given cascade has the Lorentzian lineshape. In this case, while at large $m$ the profile of the cascade at a fixed distance from its bottom converges to a certain universal form, the peak value of the modulus of the electric (magnetic) field amplitude increases with an increase in $m$. Thus, the increase in $m$ makes the resonances more pronounced. At finite dissipation the growth of the amplitude of the resonances eventually saturates, and the lineshape becomes a periodic function \mbox{of $m$.}

It is important that at large enough $m$ the positions of the electric resonances for a partial mode with multipolarity $n$ correspond to those for magnetic resonances with the multipolarity $n+1$, the electric mode with the multipolarity $n+2$, etc. The same is true for the magnetic modes.

For a given multipolarity the points of both the types of the resonances together (i.e., electric and magnetic) are situated in the $m$-axis periodically with the period equal to the half of the period for each type of the resonances separately. It means the points of a magnetic resonance are situated just in the middle of the two adjacent points of the electric resonances and vice versa.

We have shown that the cascade of the Fano resonances, exhibited with an increase in the refractive index of a scatterer, is a general, intrinsic feature of the light scattering problem. In this problem the field scattering by the same scatterer, but with the perfectly reflecting properties plays the role of a background partition, while the resonantly excited in the scatterer Mie modes correspond to the resonant ones.  The characteristics of the Fano profiles, including a simple expression for the asymmetry parameter $q$, see Eqs.~\eqref{Eq:tan_delta_PRS}, \eqref{Eq:q_and_epsilon_a}, have been obtained from ``the first principles" based upon the identical transformations of the exact Mie solution.

We have demonstrated that in the non-dissipative limit the discussed resonances (for both the inner and outer problems) possess a scale invariance, so that at any fixed value of the size parameter $x$ any resonance line in the infinite cascades of the resonances may be reduced to the universal, $m^{({\rm res})}$-independent form by the scale transformations, Eqs.~\eqref{Eq:scale_m}, \eqref{Eq:scale_cd}. It should be stressed that the universality extends to both the shape of the lines and the mutual positions of lines with different $n$ with respect to each other. The quantitative applicability conditions for the non-dissipative and dissipation-controlled regimes as well as the corresponding crossover points have been obtained in the explicit form.

The peculiarities of the resonances at a fixed refractive index and a varying size parameter, $x$ have been revealed. It has been shown that, generally speaking, the entire domain $0 \leq x < \infty$ is partitioned into three subdomains: two dissipation-controlled subdomains (at small and very large $x$) are separated by a non-dissipative one. The explicit expressions, determining the boundaries between the subdomains as well as the formulas for the lineshape and linewidth have been derived. The linewidth is minimal in the non-dissipative domain and increases in the dissipation-controlled subdomains with an increase in the departure of $x$ from the corresponding boundary.

For the size parameter lying in the non-dissipative subdomain a high concentration of the electromagnetic field within the particle may be achieved owing to individual narrow-line partial  resonances of modes with a high $Q$-factor. For the realistic values of the complex refractive index, typical for a number of common semiconductors in the visible and near IR range of the spectrum, the peak value of the density of electromagnetic energy in this case may exceed the one in the incident plane wave in the three orders of magnitude. In contrast, for the size parameter, lying in the dissipation-controlled subdomains even larger concentration may be realized due to multiple overlap of rather broad resonance lines of modes with of different types (magnetic and electric) and different multipolarity. The discussed effect of the giant concentration of the electromagnetic field within a particle with a high refractive index may have great importance in medical applications (such as, e.g., cancer therapy~\cite{El-Sayed:LMS:2008}), in the design and fabrication of high-nonlinear nanostructures, etc. A detailed study of this important issue is a separate problem and will be reported elsewhere.

\begin{acknowledgments}
The authors are grateful to B.S. Luk'yanchuk for the critical reading of the manuscript and valuable comments. The work of AEM was supported by the Australian Research Council via Future Fellowship \mbox{program (FT110100037).}\\
\end{acknowledgments}
\vspace*{-3mm}
\appendix
\section{Proof of identity, Eq.~\eqref{Eq:identity_2}}

The proof is based upon Eq.~\eqref{Eq:identity_1}. According to it and presentation of $\xi_n$ in the form \mbox{$\xi_n = \psi_n - i\chi_n$,} with real functions $\psi_n,\;\chi_n$, we may write the following chain of identities:
\begin{align}
  \frac{\xi_n(x)}{\xi_n'(x)}  \equiv  \frac{\psi_n'(x)\xi_n(x)}{\psi_n'(x)\xi_n'(x)} \equiv \frac{\psi_n(x)\xi_n'(x) -i}{\psi_n'(x)\xi_n'(x)} \tag{A.1}\label{Eq:xsi_n/xsi_n_prime}\\
  \equiv \frac{\psi_n(x)}{\psi_n'(x)} - \frac{i\xi_n'^\ast(x)}{\psi_n'(x)|\xi_n'(x)|^2}, \nonumber
\end{align}
where asterisk means complex conjugation. Then, bearing in mind the same presentation $\xi_n = \psi_n - i\chi_n$, and taking the imaginary part of Eq.~\eqref{Eq:xsi_n/xsi_n_prime} we obtain
\vspace*{-2pt}
\begin{equation*}
  {\rm Im}\frac{\xi_n(x)}{\xi_n'(x)} = -\frac{1}{|\xi_n'(x)|^2}.
\end{equation*}
\mbox{}\\
Finally, multiplying it by $\xi_n'^2(x)$ and taking modulus, we arrive at identity, Eq.~\eqref{Eq:identity_2}.

%
%
%

\bibliographystyle{aipnum4-1}
\bibliography{Mie_Fano}

\end{document}